\documentclass[11pt]{iopart}
\usepackage{iopams}
\usepackage{epsfig,graphicx,subfigure}
\newcommand{\be}{\begin{equation}}
\newcommand{\ee}{\end{equation}}

\newcommand{\BE}{\begin{eqnarray}}
\newcommand{\EE}{\end{eqnarray}}

\begin{document}
\setlength{\parindent}{0em}
\title[Anomalous fluctuations Minority Games and related multi-agent models]{Anomalous fluctuations in Minority Games and related multi-agent models of financial markets}

\author{Tobias Galla\dag\ddag, Giancarlo Mosetti\P\S, Yi-Cheng Zhang\S
}

\address{\dag\ The Abdus Salam International Center for Theoretical Physics, Strada Costiera 11, 34014 Trieste, Italy}
\address{\ddag Consiglio Nazionale delle Ricerche - Istituto Nazionale per la Fisica della Materia (CNR-INFM), Trieste-SISSA Unit, V. Beirut 2-4, 34014 Trieste, Italy}
\address{\P  Institute for Scientific Interchange ISI, Viale S. Severo 65, 10133 Torino, Italy}
\address{\S\ Department of Physics, University of Fribourg, Chemin du Musee 3, CH - 1700 Fribourg, Switzerland}
\begin{abstract}
We review the recent approaches to modelling financial markets based
on multi-agent systems. After a brief summary of the basic stylised
facts observed in real-market time-series we discuss some simple
agent-based systems which are currently used to model financial
markets. One of the most prominent examples is here the Minority Game (MG),
which we address in some more detail. After a brief discussion of its
basic setup and general phenomenology we summarise the main findings
of the statistical mechanics analysis and discuss the emergence of
stylised facts in extensions of the MG near their phase transitions between
efficient and predictable regimes. We then turn towards more
realistic variants which comprise heterogeneous populations of
agents, with different memory capabilities, different inclinations to
trade and varying expectations on the future evolution of the market.
Finally we give a short outlook on potential future work in this area.

\end{abstract}


\ead{\tt galla@ictp.trieste.it}
\ead{\tt mosetti@isi.it}
\ead{\tt yi-cheng.zhang@unifr.ch}

\section{Introduction}
The modelling of financial markets and other social systems by means
of individual-based simulations has attracted a significant amount of
attention in recent years, both in the economics and socio-economics
community, as well as among statistical physicists
\cite{Axelrod,Gilbert&Troitzsch,LLS, ChalMarsZhanBook,CoolBook,JohnJefHui03,Mantegna&Stanley}. The range of phenomena to which
this approach is applied is broad, and includes not only the modelling
of traders in financial markets, but also opinion formation and
decision dynamics, epidemic spreading, the behaviour of crowds and
pedestrians, as well as vehicular traffic and co-operative behaviour
in animal swarms \cite{helbing1,helbing2,schreck,barabasi,bak}. The
term `individual' in individual-based approaches to the modelling of
such systems may here stand for traders in a financial market, cars on
a highway or companies in an economic network. While practitioners
often make use of sophisticated models with a variety of different
types of agents, statistical physicists have become interested in
minimalist models of such phenomena, which ideally allow for
analytical solutions. While the detail and complexity of more
realistic models often impedes analytical approaches, the reduced
models considered by physicists are often tractable with methods from
statistical physics, and their behaviour can be characterised
analytically for example by exact or approximative calculation of
their phase diagrams. Furthermore the restriction to minimalist models
allows one to focus on the parameters and features of the model which
are truly responsible for its behaviour (e.g. by systematic adding or
removal of individual features), whereas the underlying mechanisms may
be obscured and thus harder to detect in more sophisticated and
detailed models. Apart from the methodology, concepts of physics often
allow one to shed light on phenomena in adjacent fields and to address
them from novel viewpoints. The concepts of self-organised criticality
\cite{bak} or replica symmetry breaking and the corresponding rugged
landscape picture of spin glass physics may here serve as examples
\cite{PaMeVi}.  Agent-based models and models from statistical physics
share common features in that in both cases one is interested in
complex co-operative behaviour on a {\em macroscopic} scale emerging
from relatively simple interactions at the {\em microscopic} level,
i.e. between the interacting agents. One may thus expect that
techniques and methods developed originally in the context of physics
may be successfully employed in other disciplines as well in order to
study complex many-agent systems and their emergent collective
phenomena.

This review focuses on some such approaches in the context of the
modelling of financial markets by agent-based trading models. Although
the statistical physics analysis is limited to minimalist models,
several such model systems have been seen to show remarkably complex
behaviour, including phase transitions, non-trivial co-operation and
global adaptation
\cite{ChalMarsZhanBook,CoolBook,JohnJefHui03,Mantegna&Stanley}. From
the point of view of the description of real-world phenomena they also
display what is referred to as `stylised facts' in the economics
literature, namely anomalous non-Gaussian behaviour and fluctuations
in their global observables, similar to statistical features seen in
real market data. This positions the agent-based models considered by
statistical physicists right at the boundary between being
analytically solvable and at the same time sufficiently realistic, and
constitutes some of the appeal of such models.

The objective of this chapter is to review some of the developments in
the study of agent-based models, stressing the emergence of anomalous
behaviour. While some, potentially non-representative focus is given
to Minority Game market models and the authors' own work, we also
present a brief discussion of some other related agent-based models.

\section{Overview and stylised facts in financial markets}
\subsection{Introduction}

`Stylised facts' in the context of finance is a term used by
theoretical economists and by practitioners to summarise statistical
features of time series generated by financial markets. Most notably
financial markets generate non-Gaussian time series with power-law
characteristics, long-range correlations in time, scaling behaviour
and anomalous fluctuations
\cite{Mantegna&Stanley,Bouchaud&Potters}. Non-Gaussian behaviour is
indeed found in many different contexts, already more than 100 years
ago, Vilfredo Pareto \cite{Pareto} pointed out that in numerous
countries and at various points in time the wealth distribution
follows a power-law, i.e. $N\sim w^{-\alpha}$, where $N$ is the number
of people having wealth greater than $w$. Over the past decades
similar power-law behaviour has been detected in numerous other
phenomena, ranging from financial time series, to sand piles, the
distribution of word frequency in texts, citations of scientific
papers, hits on web pages, firm sizes, populations of cities so forth
\cite{bak,newman}.

Anomalous, non-Gaussian fluctuations are not only observed in the
above systems, apparently unrelated to physics, but also in systems of
condensed matter physics, and in models of statistical mechanics.
Anomalous behaviour is here found in a variety of contexts, from
magnetic materials, over solutions of polymers, gels, glasses, fluid-
and superfluid helium, water at its boiling point, and even in
experiments of elementary particle physics and in cosmology
\cite{Kadanoff,Stanley}. The common feature of these systems is that they display phase transitions, i.e. depending on external control parameters (such as
temperature or pressure) the behaviour of one given system can be of
different types. Not only do the characteristics of these systems
change drastically at their transition points, but thermal
fluctuations attain non-Gaussian features, very similar to those
observed in unphysical systems mentioned above. Close to their
critical points systems of condensed matter physics display long-range
correlations, with an algebraic as opposed to an exponential decay of
spatial and temporal correlations, leading to large-scale collective
behaviour, ergodicity breaking and other so-called `anomalous'
characteristics. Power-law correlation functions are here a reflection
of scale-free behaviour of the system, i.e. of fractal properties and
self-similarity. This invariance under re-scaling of units has led to
the development of powerful methods of statistical physics, most
notably of renormalisation group, which allow to address critical
phenomena efficiently \cite{Kadanoff,Stanley,Cardy}.

\subsection{Anomalous fluctuations in financial markets}

\begin{figure}[t]
\begin{center}
\vspace{10mm}
\includegraphics[width=5cm,angle=270]{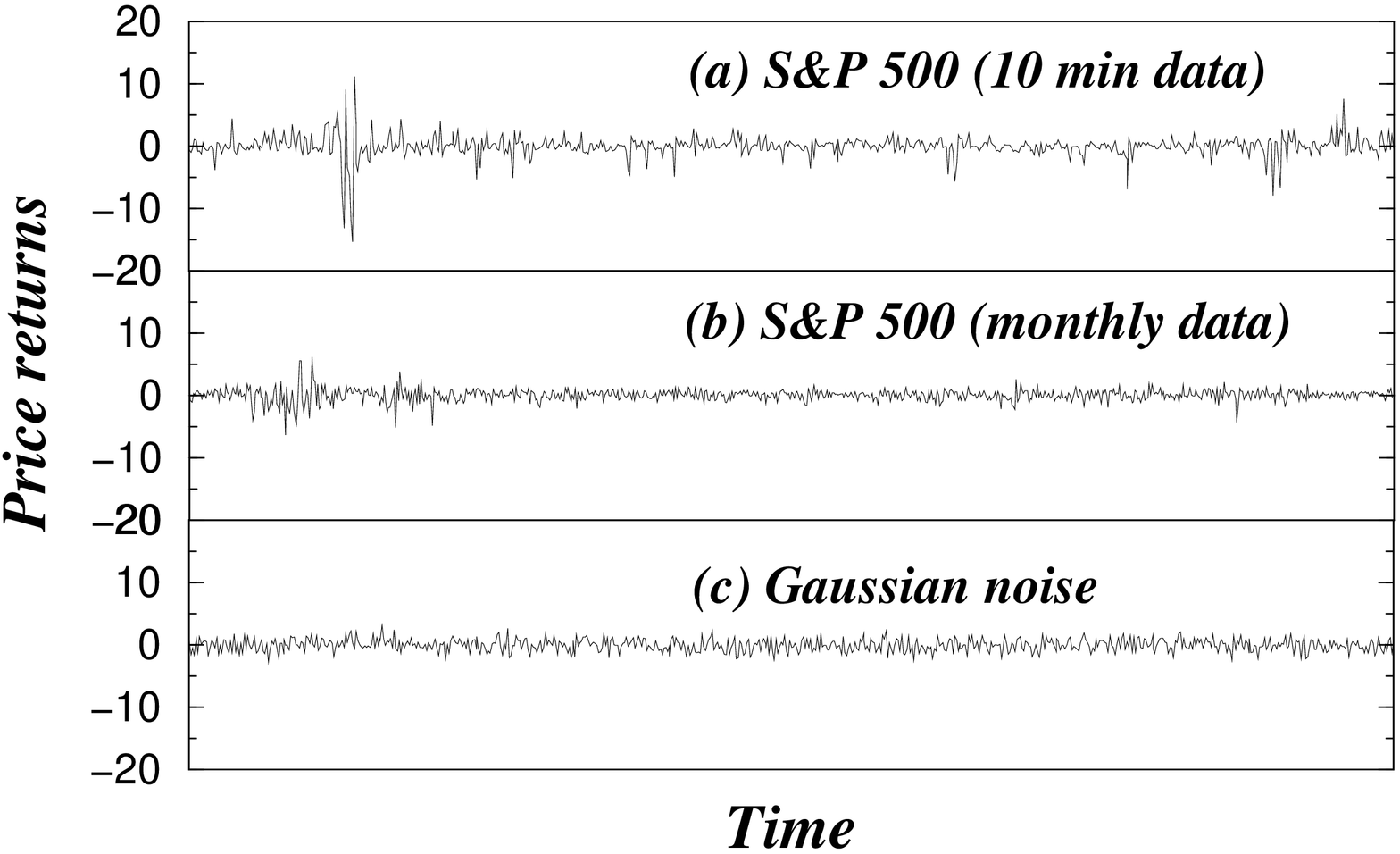} ~~~
\includegraphics[width=5cm,angle=270]{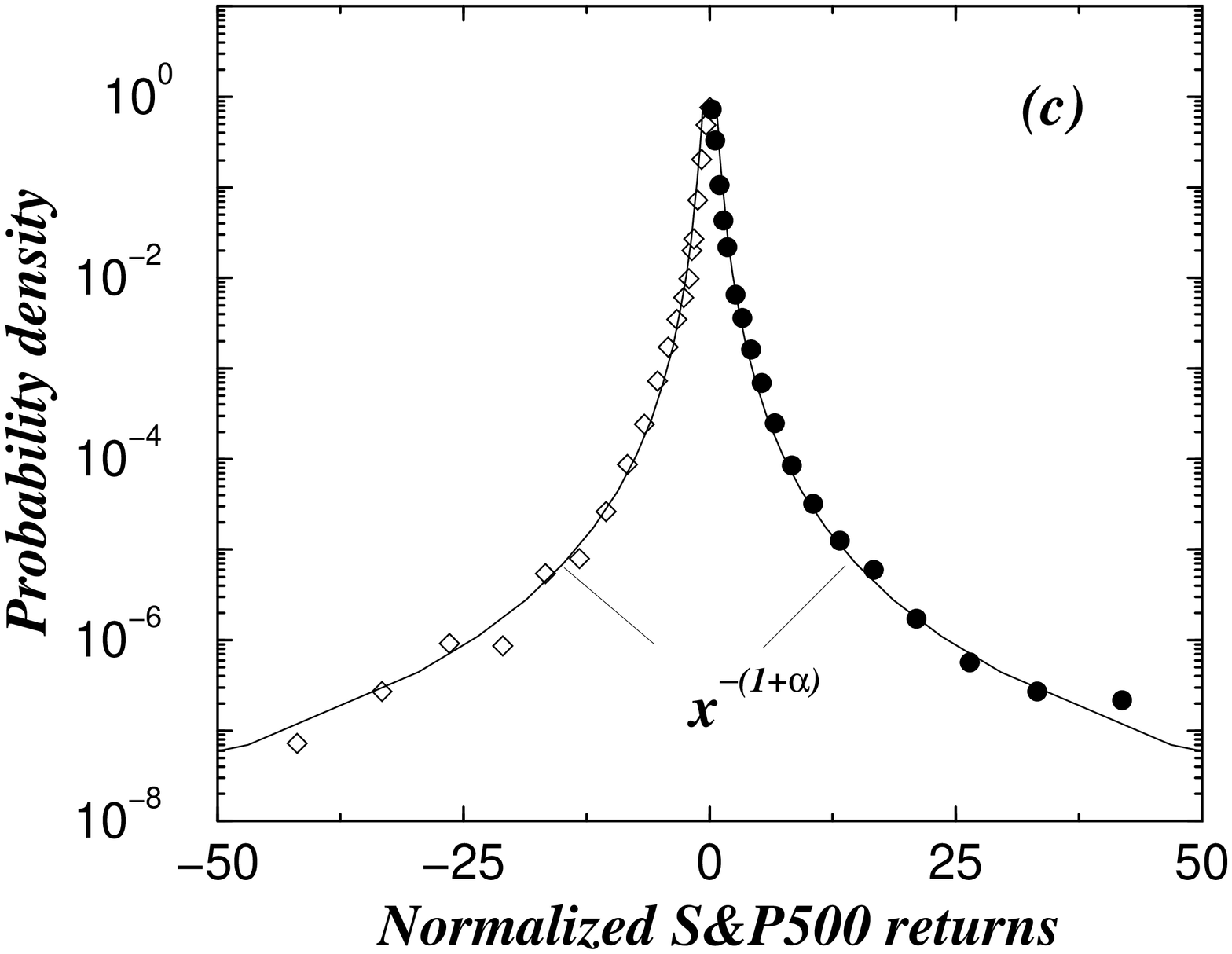} 
\end{center}

\caption{{\bf Left:} Sequence of 10$\,$min returns of the S\&P 500 index (top), 1$\,$month returns (middle), and a realisation of a Gaussian random walk for comparison. {\bf Right:} Linear-log plots of the probability distribution for the normalized S\&P500 returns. The solid lines
are power-law fits with exponents $1+ \alpha \approx 4$. Figures taken
from \cite{Gop&al}.\label{fig:sptimeseries}}
\end{figure}

Power-law statistics and non-Gaussian behaviour have been detected in
a variety of different markets, and are mostly independent of the
specific trading rules or external circumstances at the given market
place
\cite{Mantegna&Stanley}. For example, if we denote  the
price of an asset or an index by $X(t)$, then the log-return at time
scale $\Delta_t$ is defined as
$r_t=\ln\left(X(t)/X(t-\Delta_t)\right)$.  Empirical studies show that
the distribution of log-returns has power-law tails with exponents
between $3$ and $5$ \footnote{This result holds for different time
scales $\Delta_t$.}\cite{Gop&al}.  Another interesting feature which
is observed in financial markets is the so-called {\em volatility
clustering}: while the correlation of returns decays relatively
quickly in time, the correlation of the absolute log-returns, defined
as $C_{|r|}(\tau)=\left\langle |r(t+\tau)||r(t)|\right\rangle $,
follows a long-range power-law distribution with an exponent ranging
from $0.1$ to $0.3$ \cite{Lo,Liu&al}. Non-Gaussian distributions are
also observed when looking at the trading volume or the number of
trades per time. We illustrate these observations in
Fig. \ref{fig:sptimeseries}, where a financial time-series with a
characteristic non-Gaussian behaviour is shown, along with its
algebraically decaying return distribution.

Given the abundance of financial data physicists have become involved
in the analysis of time series of markets, as well as in research
aiming to model the processes generating these data. This field at the
borderline of statistical physics, econometrics and financial
mathematics is now known as {\em econophysics} \cite{chalweb}.  The attempts to model
financial markets here roughly fall into two classes, phenomenological
models and agent-based approaches. In the next section we will briefly
discuss both, and then concentrate on the latter agent-based approach,
as mentioned in the introduction. We will here mostly focus on qualitative aspects, and will not enter mathematical details of the various models. For a recent textbook on mathematical aspects of MGs and related models see also \cite{CoolBook}.

\section{The agent-based approach}
\subsection{Random walks and other phenomenological models}
Many approaches to anomalous statistical properties of financial data
are based on direct modelling of stock market indices as stochastic
processes. Such models do not deal with the market on a microscopic
level, i.e. with individual agents, but with aggregate quantities
only, such as the stock market indices, their return statistics and
fluctuations on a macroscopic level. The very first of such theories
dates back to Bachelier's `Th\'eorie de la Sp\'eculation' from 1900
\cite{bachelier} and describes the stochastic evolution of market
indices as simple Gaussian random walks. While it is now well
established that market data is fundamentally non-Gaussian,
Bachelier's theory of financial markets as log-normal random walks
still forms the basis of many phenomenological approaches to
finance. Most notably, the Black-Scholes theory \cite{BS} commonly
used for the pricing of financial derivatives relies entirely on
Gaussian assumptions, and hence on Bachelier's approach.

More realistic approaches to finance with stochastic processes are of
course no longer based on purely Gaussian processes, but use L\'evy
flights and related processes to model the temporal evolution of
financial indicators \cite{Bouchaud&Potters}. Many different classes
of stochastic processes have here been used, partially relying on the
fact that multiplicative as opposed to additive Gaussian noise
directly leads to power-law behaviour.

\subsection{Agent-based models}

While phenomenological models can describe many features observed in
real markets, they are at least to some extent not fully satisfactory
in that they do not allow for a derivation of global quantities from
the most basic level of description, namely the direct interaction
between agents. These phenomenological and agent-based approaches are,
to some degree, analogous to the description of physical phenomena
through thermodynamics and statistical physics respectively.  While
the laws of thermodynamics, such as for example the phenomenological
equation describing an ideal gas, find their justification mostly in
the purely empirical observation, only a direct kinetic theory allows
for a description on the level of individual molecules. During 19th
century such a kinetic theory was being developed, starting from
Newton's laws and using probabilistic arguments, and results from
thermodynamics were re-derived from first principles, marking the
beginning of statistical mechanics.

One can think of phenomenological models of financial markets as the
analogue of the description of physical phenomena on the level of
thermodynamics.  Empirical laws like the ideal gas equation here
correspond to the stylised facts empirically found in time series of
financial markets.  The aim of phenomenological models is thus, in
essence, to devise and study stochastic processes for macroscopic
observables, without considering the detailed microscopic  interaction which bring about the global phenomena.

The counterpart of statistical mechanics would then be given by models
which start from the level of the individual interacting particles of
a financial market, i.e. the traders. While in physics particles can
be considered the basic elements, in markets this role is played by
financial agents.  The objective of statistical mechanics approaches
to financial markets is thus to study the interactions between
individual agents, and to derive equations describing global
quantities from these microscopic rules of engagement. Naturally the
analogy between particles and agents is limited, for example because
agents are intelligent and particles are of course not. However,
concepts and techniques from physics can be transferred to the study
of agent-based systems, which mathematically turn out to be
surprisingly similar to models of statistical physics.

The remainder of this review will focus on agent-based models. These
allow one to address fundamental questions for example as to where
anomalous fluctuations may actually come from on the level of the
behaviour of the agents. Agent-based models of financial markets have
attracted substantial attraction, both in the economics and in the
physics communities. They easily accessible by computer simulations,
so that effects of changes in the rules of engagement of the agents,
the ecology of the model-market or variations in external control
parameters are readily examined. The role of physicists has here
mostly been to devise simple minimalist models of financial markets,
which are also accessible analytically and conceptually by tools of
statistical mechanics.

\section{Minority Games and related models}
\subsection{Introduction}
Neoclassical economics usually assumes that agents have perfect, logic
and deductive rationality, as well as full information about the
market \cite{walras,merton,samuelson}. This implies, among other
things, that they are homogeneous and that they act always in the best
possible way. Common sense suggests that this view is quite far from
reality and for different reasons people are neither completely
rational nor deductive and homogeneous. Starting from this
observation, leading economist Brian Arthur introduced a very simple
model, now known as the `El Farol bar problem', to illustrate bounded
rationality and inductive reasoning. In Arthur's model agents have
only limited information and modes of reacting to it at hand, and
learn inductively from past experience. The mathematical abstraction
of this model resulted in the Minority Game (MG) \cite{ChalZhan97},
which now also serves as one of the most popular toy models for
financial markets.

Despite its simplicity, the MG exhibits a very rich behaviour, with
phases in which the resulting model-market behaves fully inefficiently
and others in which the time-series generated by the market still has
some information-content, leading to an effective predictability. As a
function of the model parameters, the agents can either co-operate
successfully and achieve total gains higher than if they were to play
at random, or in other regimes fail to adapt successfully, resulting
in high losses.

In its original formulation the statistics of the MG are mostly
Gaussian (if large populations of agents are considered), and there
are no signs of volatility clustering and other stylised facts in the
time-series generated by the most basic MGs. Small modifications,
however, can produce the fat tailed distributions similar to those
observed in financial markets.  Apart from the appeal of the MG as a market model, it has also attracted a significant interest as a model of statistical mechanics, and the methods of theoretical spin glass physics have been seen to provide insight into its phase behaviour, and have revealed new types of global co-operation, complexity and phase transitions, previously unknown in models of disordered systems theory \cite{ChalMarsZhanBook,CoolBook}.

\subsection{Definition of the MG model}
The MG is a mathematical abstraction of the El-Farol bar problem. In
the latter a group of say $100$ agents have to decide whether or not
to attend a bar at a given night every week. In general people will
only enjoy being at the bar if not more than say $60$ people attend
that night, as otherwise the bar will be too crowded. Every agent has
to decide independently every week whether to go or not, and has to
make that decision on the basis of the time-series of previous
attendances. El-Farol agents in Arthur's model are agents of bounded
rationality who learn inductively. Each of them holds a set of
predictors, mapping the past attendances onto a future (predicted) one,
and based on these idiosyncratic predictions the individual agents
decide on whether or not to attend the bar.

The basic MG model describes this systems as a population of $N$
agents (with $N$ an odd integer), in which every player $i$ has to
take a binary decision $b_i(t)\in\{-1,1\}$ at every time-step $t$. The
total attendance is then computed as

\begin{equation}
A(t)=\sum_{i=1}^N b_i(t),
\end{equation}

and agents who take the minority decision are rewarded after each
round, that is if the number of agents playing $1$ is higher than the
number of those playing $-1$ then all $-1$-players are rewarded and
vice versa. In order to take their decisions agents are each equipped
with a pool of $S$ so-called `strategy tables'. Assuming that
agents can remember the correct minority decisions of the last $M$
rounds, that is they can resolve $P=2^M$ different histories of the
game, a strategy $a_{is}$ provides a map from all possible $M$-step
histories onto the binary set $\{-1,1\}$, i. e.
\be
a_{is}:\{1,2...,P\}\rightarrow \{-1,1\}.
\ee

In order to take their trading decisions agents follow an inductive
learning dynamics. They are limited to the strategies assigned to them
at the beginning of the game, reflecting the boundedness of their
rationality. These strategy assignments are typically generated at
random, and result in a heterogeneous population of traders. At each
time step each agent has to choose which of his strategy tables to
use, and the agents do so by learning from past experience. A score
$U_{i,s}$ (utility function in terms of economics) is assigned to each
strategy table and updated at each step of the game as follows:
\be\label{eq:mgupdate}
U_{i,s}(t+1)=U_{i,s}(t)-a_{i,s}^{\mu(t)}A(t).
\ee
Here $a_{i,s}^{\mu(t)}$ is the action prescribed by the strategy $s$
of agent $i$ when the history $\mu(t)$ occurs. The term
$-a_{i,s}^{\mu(t)}A(t)$ reflects the minority rule: when
$a_{i,s}^{\mu(t)}$ and $A(t)$ have opposite signs, i.e. if the strategy
predicted the correct minority decision, then the strategy score
$U_{i,s}$ is increased. If $a_{i,s}^{\mu(t)}A(t)>0$, the score is
decreased. Each agent then at each time step uses the strategy table
in his pool of strategies with the highest score, i.e. the one that
would have performed best so far, had this particular player always
used the same strategy. In case of equal scores between some
strategies, the agent chooses between those at random.

\subsection{Motivation as market model and price process}
The MG can be understood as a simplified emulation of a financial
market. The binary decisions of the agents are interpreted as `buying'
versus `selling', and hence the aggregate action corresponds to an
`excess demand' in this model market, and leads to movements of the
(logarithmic) price. More precisely, if $p(t)$ is the price of the
good traded on the MG-market, then $\log p(t+1)=\log
p(t)+\frac{A(t)}{\lambda}$, with $A(t)$ the above aggregate action
$A(t)=\sum_{i=1}^N b_i(t)$. $\lambda$ here reflects the liquidity of
the market and represents a proportionality constant. Traders in the
MG take their decisions based on commonly available information, which
may be related to the past price history of the market under
consideration or to information provided externally. The payoff
structure of the MG can here be derived from expectations of the
agents on future price evolution \cite{Marsili}. Minority game agents
effectively correspond to contrarian traders who expect the price to
go down in step $t+1$ if it went up in step $t$. It is straightforward
to extend the model to majority game or mixed minority/majority games
(in which some agents play an minority game, and others a majority
game)
\cite{DeMaGiarMose03,CoolBook}. Majority game agents here
represent so-called `trend-followers' in a real market, their
expectation is that the price movement of time step $t+1$ will be in
the same direction as at time $t$, so that it is favourable to buy
when the majority of players is buying, and to sell whenever the
majority is selling.

\subsection{Behaviour and main feature of the MG}
We will now turn to a brief discussion of the main phenomenology of
the MG. While it follows directly from symmetry arguments that the
time average of total bid vanishes for large system sizes
(i.e. $<A(t)>=0$), the fluctuations of $A(t)$ display a remarkably
complex behaviour when looked at as a function of the memory
capacities of the agents. It was observed in \cite{SaviManuRiol99}
that the ratio $\alpha=P/N$ is the relevant scaling parameter of the
model, i.e. that $\sigma^2/N=\overline{<A(t)^2>}/N$ \footnote{Here
$<\dots>$ stands for a temporal average, while $\overline{\cdots}$ is
the notation used for an average over the quenched disorder, i.e the
strategy assignments.} and other observables depend on the memory
length $M=\log_2 P$ and the system size only through the ratio
$\alpha=P/N$. The variance $\sigma^2$ of $A(t)$, also referred to as
the {\em volatility} is a measure for the efficiency of the game in
terms of co-operation and global gain. The smaller $\sigma^2$ the
smaller is the group of losing agents (i.e. the majority group), with
$A(t)\approx 0$ (i.e. $\sigma^2 \approx 0$) corresponding to a
situation in which close to half of the agents take either trading
decision, i.e. with roughly $N/2$ agents losing and the other $N/2$
agents winners (up to corrections to lower order in $N$). A detailed
analysis shows that the global gains are in fact given by $-\sigma^2$,
due to the inherent negative-sum character of the MG global gains are
always non-positive.

\begin{figure}[t]
\begin{center}
\includegraphics[width=14pc]{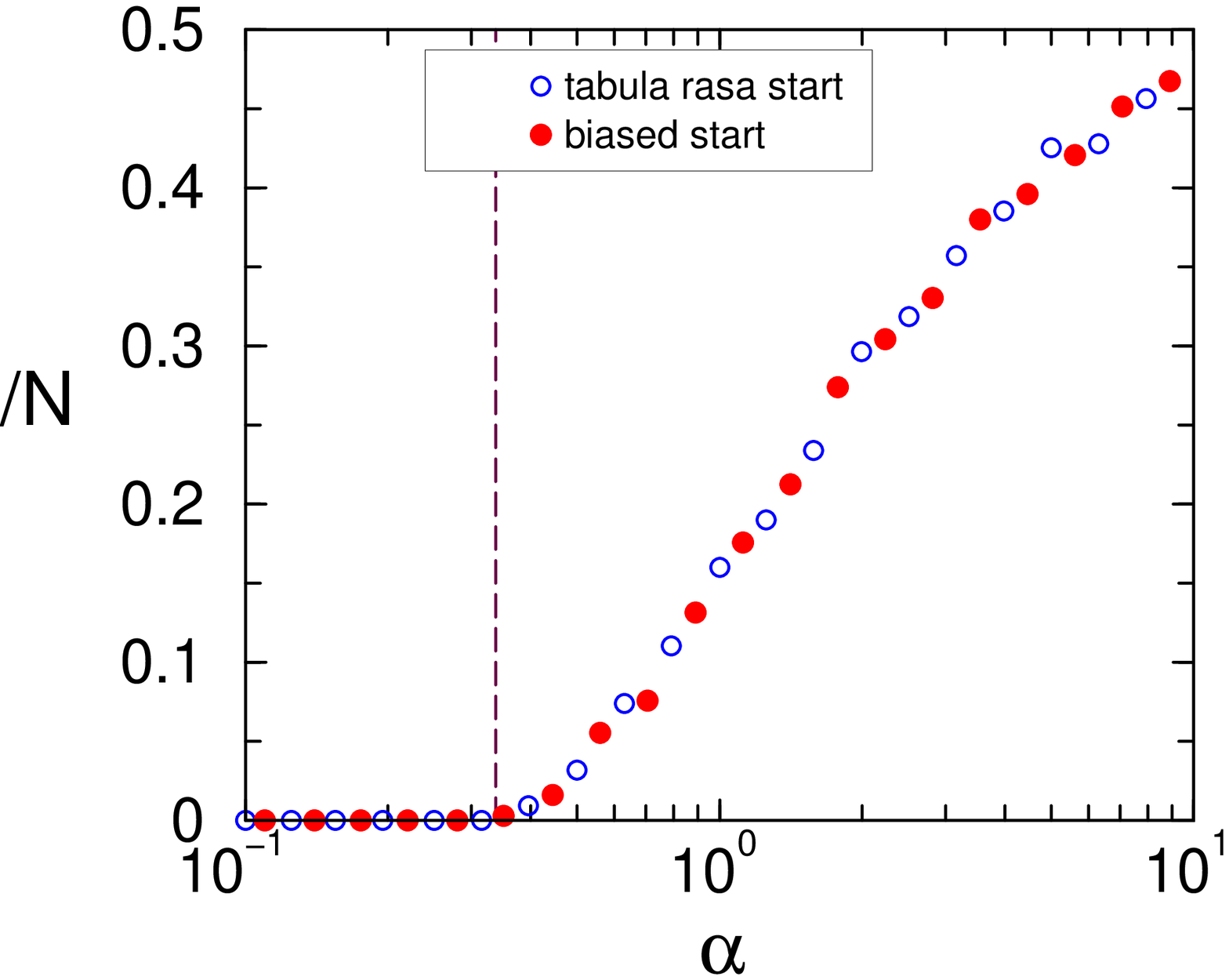}~~~~~~~~~~~~\includegraphics[width=14pc]{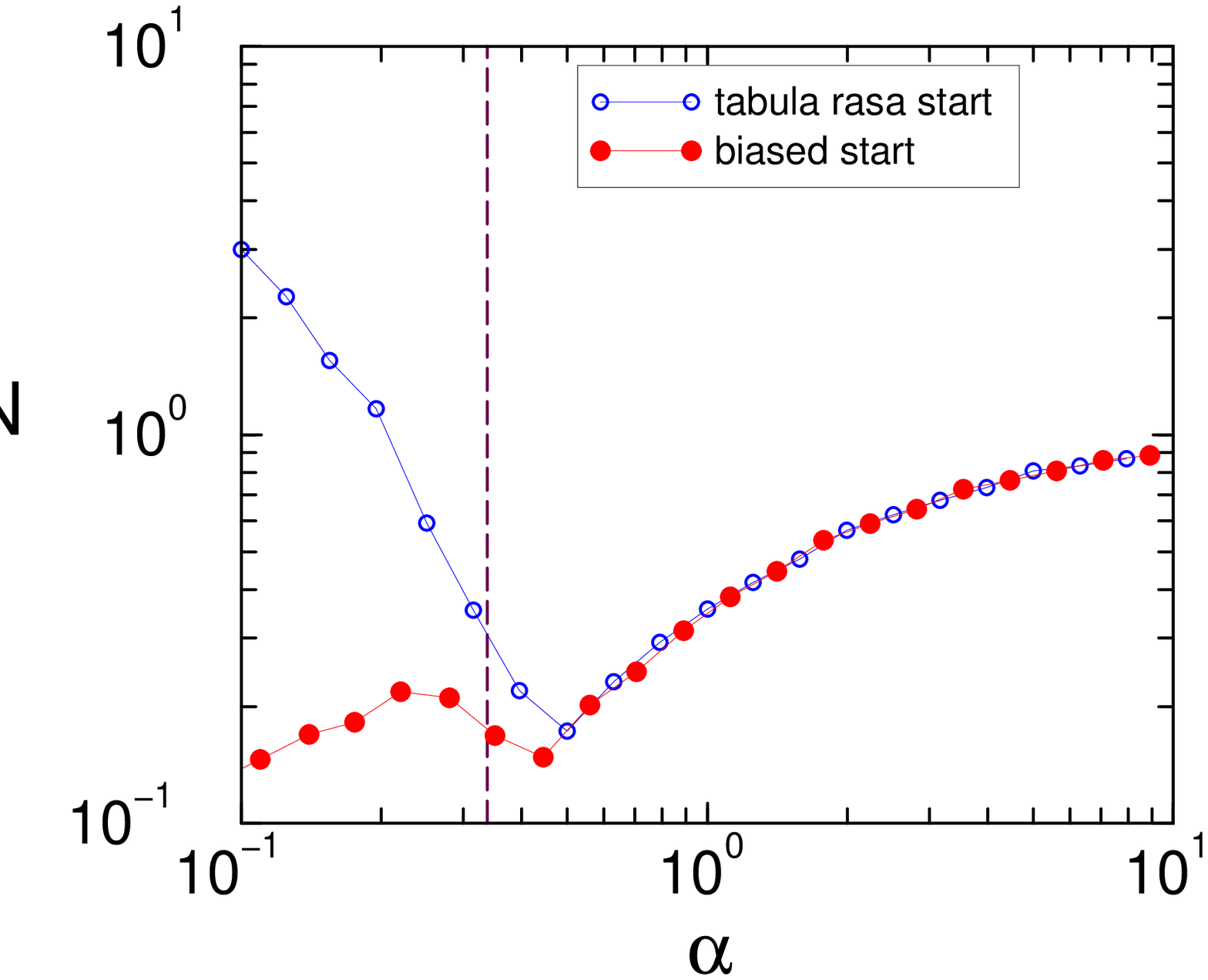}
\end{center}
\caption{(Colour on-line) {\bf Left:} Predictability as a function of $\alpha$ for the standard MG for {\em tabula rasa} starts (open markers) and biased initial conditions (solid markers). {\bf Right:} Volatility.}
\label{fig:basic}
\end{figure}

The behaviour of the re-scaled volatility $\sigma^2/N$ as a function
of $\alpha$ is depicted in Fig. \ref{fig:basic}. For large $\alpha$
one finds $\sigma^2/N\approx 1$ which corresponds to the so-called
`random trading limit'. As $\alpha$ is decreased one finds a minimum
of $\sigma^2/N$ at intermediate $\alpha$, and a large volatility
$\sigma^2/N>1$ as $\alpha\to 0$ (for zero initial conditions). In this
regime the co-operation among MG agents is worse than the random
trading limit, mostly due to collective motion of whole crowds of
agents taking the same trading decision (and of the corresponding
anti-crowds) \cite{JohnJefHui03,Jefferies&al,Johnson&al}.

Further insight can be obtained by defining the quantity 
\begin{equation}
H=\sum_{\mu=1}^P \left\langle A|\mu \right\rangle^2,
\end{equation}
where the $<A|\mu>$ is the average of $A$ conditional to the occurrence
of a particular history string $\mu$.  $H$ is here a measure for the
predictability of the system: if $<A|\mu>\neq 0 $ for at least one
$\mu$ then the system is predictable (in a probabilistic sense). $H=0$ thus corresponds to a fully information-efficient market, in which no exploitable information is contained in the market's time series. If $H>0$, however, statistical predictability is present, and the market is not operating at full information efficiency.

Plotting $H/N$ as a function of $\alpha$ (see Fig. \ref{fig:basic})
one finds that the value $\alpha_c$ marking the minimum of
$\sigma^2/N$ corresponds to a transition point between a predictable
phase at $\alpha>\alpha_c$ and a fully information-efficient one at
$\alpha<\alpha_c$.

At the same time, $\alpha_c$ also marks an ergodic/non-ergodic phase
transition. This can be demonstrated upon plotting the volatility
$\sigma^2/N$ as a function of $\alpha$ obtained from simulations of
the MG learning dynamics started from different initial
conditions. One here distinguishes so-called `tabula rasa' initial
conditions, for which all strategy scores are set to zero at the
beginning of the game, and so-called `biased starts', where some
strategies receive random non-zero score valuations $U_{is}(t=0)=\pm
u_0$ at the beginning (with $u_0\sim{\cal O}(1)$). A corresponding
plot is shown in Fig. \ref{fig:basic}, and demonstrates that the
system is insensitive to initial conditions above $\alpha_c$, but that
the starting point becomes relevant in the information-efficient phase
below the phase transition.
\subsection{The statistical mechanics approach}
Since its mathematical formulation in 1997 \cite{ChalZhan97} the MG
has attracted significant attention in the statistical physics
community. In particular it has been found that the MG is accessible
by the tools of the theory of disordered systems and spin-glasses
\cite{PaMeVi,FiHe}, and that its phase diagram and key observables can
be computed either exactly or in good approximation by methods of
equilibrium and non-equilibrium statistical mechanics. The analogy to
disordered systems, which have studied in the spin-glass and neural
networks communities for $30$ years, is here threefold: firstly, the
random strategy assignments which are drawn at the start of the MG and
through which the agents interact with each other correspond to
`quenched disorder' in statistical physics, that is frozen, fixed
interactions between the microscopic degrees of freedom. In spin
glasses the interactions are specified by the couplings between
magnetic spins, in neural networks they are reflected by fixed
synaptic structures governing the firing patterns of the network of
neurons. Secondly, the MG displays global frustration (not everybody
can win), similar to spin-glass models where not all interactions
between spins can be satisfied. Thirdly, in the MG every agent
interacts with everybody else (through the observation of the
aggregate action $A(t)=\sum_i b_i(t)$ which is used to update strategy
scores). This makes the MG a `mean-field' model in the language of
physics, and tools to address such systems are readily available.

We will here not enter the details of the mathematical analysis of the
MG, but would only like to mention briefly that it can be addressed by both
static methods such as replica techniques \cite{PaMeVi}, and by
dynamical approaches relying on path-integrals and dynamical
generating functionals \cite{CoolBook,HeimCool01}. We will briefly sketch the ideas of both approaches in the following two subsections.

\subsubsection{Statics: replica theory}
The starting point of the static replica approach to the standard MG
is the function $H$ as defined above \cite{ChalMarsZhanBook}. Although
$H$ is not an exact Lyapunov function of the MG learning dynamics, $H$
can be seen as a `pseudo-Hamiltonian', with the ergodic stationary
states of the MG corresponding to extrema of $H$. If one restricts the
discussion to a two-strategy MG (that is each agent $i$ holds only two
strategy tables $a_{is}$ with $s=\pm 1$), then $H$ turns out to be
given by
\be
H=\frac{1}{\alpha N}\sum_{\mu=1}^{\alpha N} \left(\sum_i\{\omega_i^\mu+\xi_i^\mu m_i\}\right)^2
\ee
where $\xi_i^\mu=\frac{1}{2}(a_{i,1}^\mu-a_{i,-1}^\mu)$ and
$\omega_i^\mu=\frac{1}{2}(a_{i,1}^\mu+a_{i,-1}^\mu)$. The $\{m_i\}$
are here soft spin variables $m_i\in[-1,1]$ and (up to normalisation)
correspond to the relative frequencies with which each player employs
each of his two strategies.

$H$ as defined above is a random function in that it depends on the
$\{\xi_i^\mu,\omega_i^\mu\}$, which reflect the random strategy
assignments at the beginning of the game. In the language of
spin-glass physics, $H$ contains quenched disorder. Information about
the stationary states of the MG can then be obtained upon minimising
$H$ with respect to the microscopic degrees of freedom $\{m_i\}$. The
starting point is the partition function
\be
Z=\int_{-1}^{1}dm_1\cdots dm_N \exp\left(-\beta H(m_1,\dots,m_N)\right)
\ee 
where the `annealing temperature' $T=1/\beta$ is taken to zero
eventually to obtain the minima of $H$, 
\be
\mbox{min}~ H = -\lim_{\beta\to \infty}\frac{1}{\beta}\ln Z.
\ee
The explicit analytical computation of the partition function for any
particular assignment of the strategy vectors this an insurmountable
task, due to complex interaction and global frustration. One hence
restricts to an effective average $\overline{\cdots}$ over the disorder,
and considers typical minima of $H$
\be
\overline{\mbox{min}~ H} = -\lim_{\beta\to 0}\frac{1}{\beta}\overline{\ln Z}.
\ee
The disorder-average of the logarithm of the partition function can here be obtained using a replica-approach, as it is standard in spin-glass physics. In particular one has
\be
f = -\lim_{n\to 0}\lim_{\beta\to 0}\lim_{N\to\infty}\frac{1}{\beta N n} \left(\overline{Z^n}-1\right)
\ee
for the free energy density in the thermodynamic limit. $Z^n$ here
corresponds to an $n$-fold replicated system, the disorder-average
generates an effective coupling between the different replicas.

The further mathematical steps are tedious, but straightforward, and
we will not report details here
\cite{ChalMarsZhanBook,CoolBook,MarsChalZecc00}, but will only mention that order
parameters such as $\lim_{N\to\infty}H/N$ and
$\lim_{N\to\infty}\sigma^2/N$ in the ergodic phase can be obtained
exactly or in good approximation from this approach. The location of
the phase transition at $\alpha_c=0.3374...$ can be identified exactly
as the point at which the so-called `static susceptibility' diverges,
indicating the breakdown of the ergodic replica-symmetric theory.

\subsubsection{Dynamics: generating functional analysis}
The dynamical approach based on generating functionals deals directly with the learning dynamics of the agents, which can be formulated as follows
\be\label{eq:udiff}
q_i(t+1)=q_i(t)-\frac{1}{N}\xi_i^{\mu(t)}\sum_j\left\{\omega_j^{\mu(t)}+\xi_j^{\mu(t)}\mbox{sgn}[q_j(t)]\right\},
\ee
if $q_i(t)$ denotes the score difference
$q_i(t)=\frac{1}{2}\left(u_{i,1}(t)-u_{i,-1}(t)\right)$ of player
$i$'s two strategies at time-step $t$ in a two-strategy MG (a generalisation to $S>2$ has recently been reported in \cite{nima}). Note that
only this difference is relevant for the agent's decision which of his
two look-up tables to use, and that at $t$ he will choose the strategy
with the higher-score, i.e. strategy table $-1$ if $q_i(t)<0$ and
strategy table $+1$ if $q_i(t)>0$. This may be summarised as him
playing strategy $\mbox{sgn}[q_i(t)]\in\{-1,1\}$ at $t$, and hence
taking trading action
$\omega_i^{\mu(t)}+\xi_i^{\mu(t)}\mbox{sgn}[q_i(t)]$
($=a_{i,+1}^{\mu(t)}$ if $q_i(t)>0$ and $=a_{i,-1}^{\mu(t)}$ if
$q_i(t)<0$).
 
One the defines a dynamical analogue of the partition function as the generating functional \cite{HeimCool01,CoolBook,CoolHeimSher01}
\be
\hspace{-2cm}Z[\psi]=\int \left[\prod_{i,t} Dq_i(t)\right] \delta(\mbox{equations of motion})\exp\left(i\sum_{it}\psi_i(t)\mbox{sgn}[q_i(t)]\right),
\ee
where `equations of motion' is an abbreviation for
Eqs. (\ref{eq:udiff}) so that the delta-functions restrict this path
integral to all trajectories of the system allowed by the update
rules. $\psi_i(t)$ is a source term, and allows one to compute correlation
functions upon differentiation with respect to the
$\{\psi_i(t)\}$. Mathematically speaking $Z[\psi]$ is the Fourier
transform of the probability measure in the space of dynamical paths
corresponding to the MG update rules, and contains all relevant
information on the dynamics of the system. Similarly to the replica
approach the evaluation of $Z[\psi]$ is intractable for individual
realisations of the disorder, but is carried out as an average over
all possible strategy assignments. This leads to a set of closed, but
complicated equations for the dynamical order parameters of the
problem (the response and correlation functions), from which one can
proceed to compute stationary order parameters such as the volatility
or the predictability. We will not report these calculations here, but
will only mention that both approaches, the static replica theory and
the dynamical generating functional analysis, lead to identical
expressions for the characteristic observables, and ultimately deliver
the same value for the phase transition point $\alpha_c$
\cite{CoolBook}.

\vspace{1em}

While there are still some mathematical subtleties to be resolved, the
MG is now essentially considered to be solved as its phase diagram
and stationary states can be computed exactly, and as valid and
convincing approximations for the volatility are available at least in
the ergodic phase. Open questions mostly relate to the behaviour in
the non-ergodic phase, where the market is information-efficient
($H=0$). Up to now, no analytical solutions have been found here.

\section{Anomalous fluctuations in Minority Games and other agent-based models}
\subsection{General remarks}\begin{figure}[t]
\begin{center}
\vspace{10mm}
\includegraphics[width=5cm]{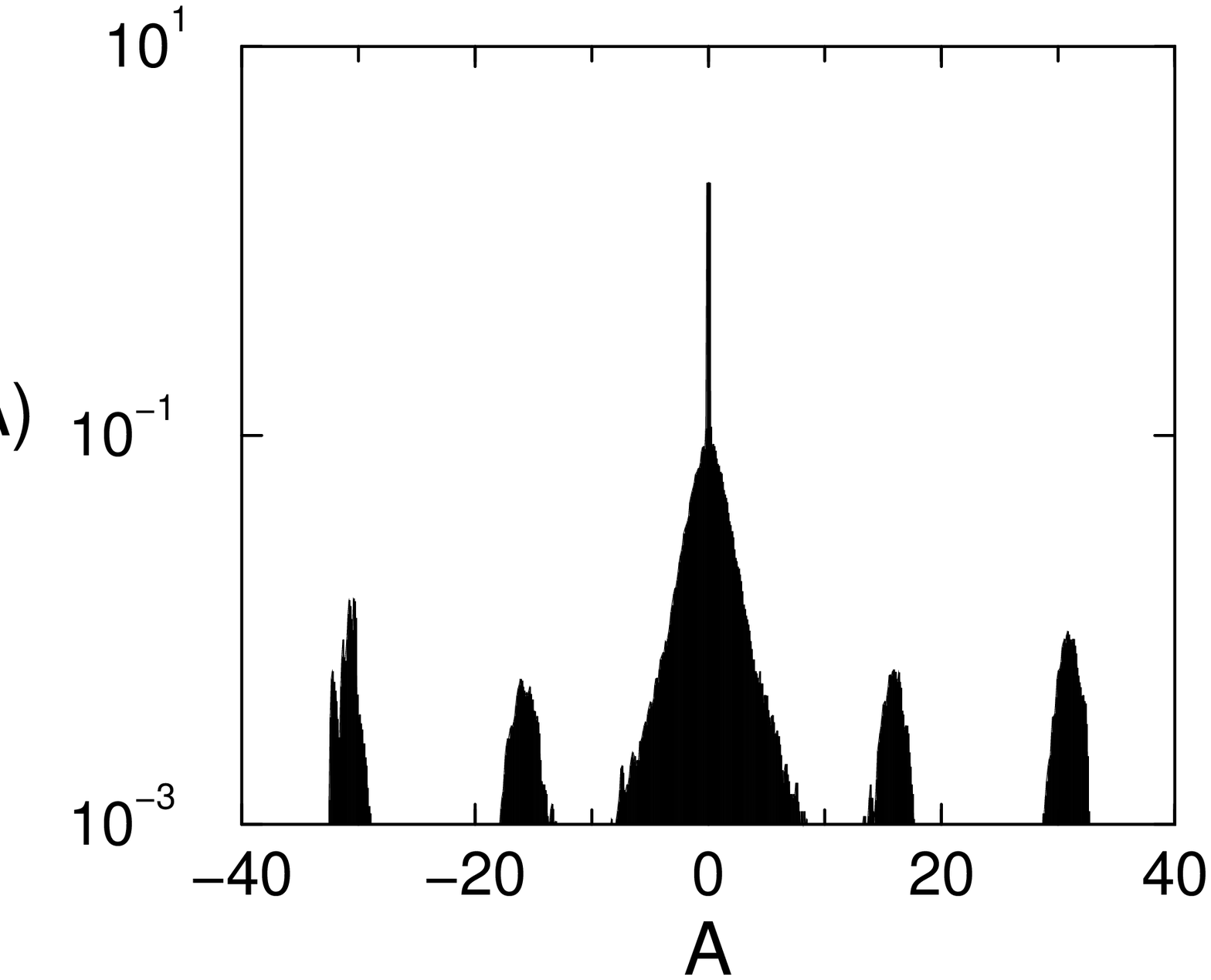} ~~~~~~
\includegraphics[width=5cm]{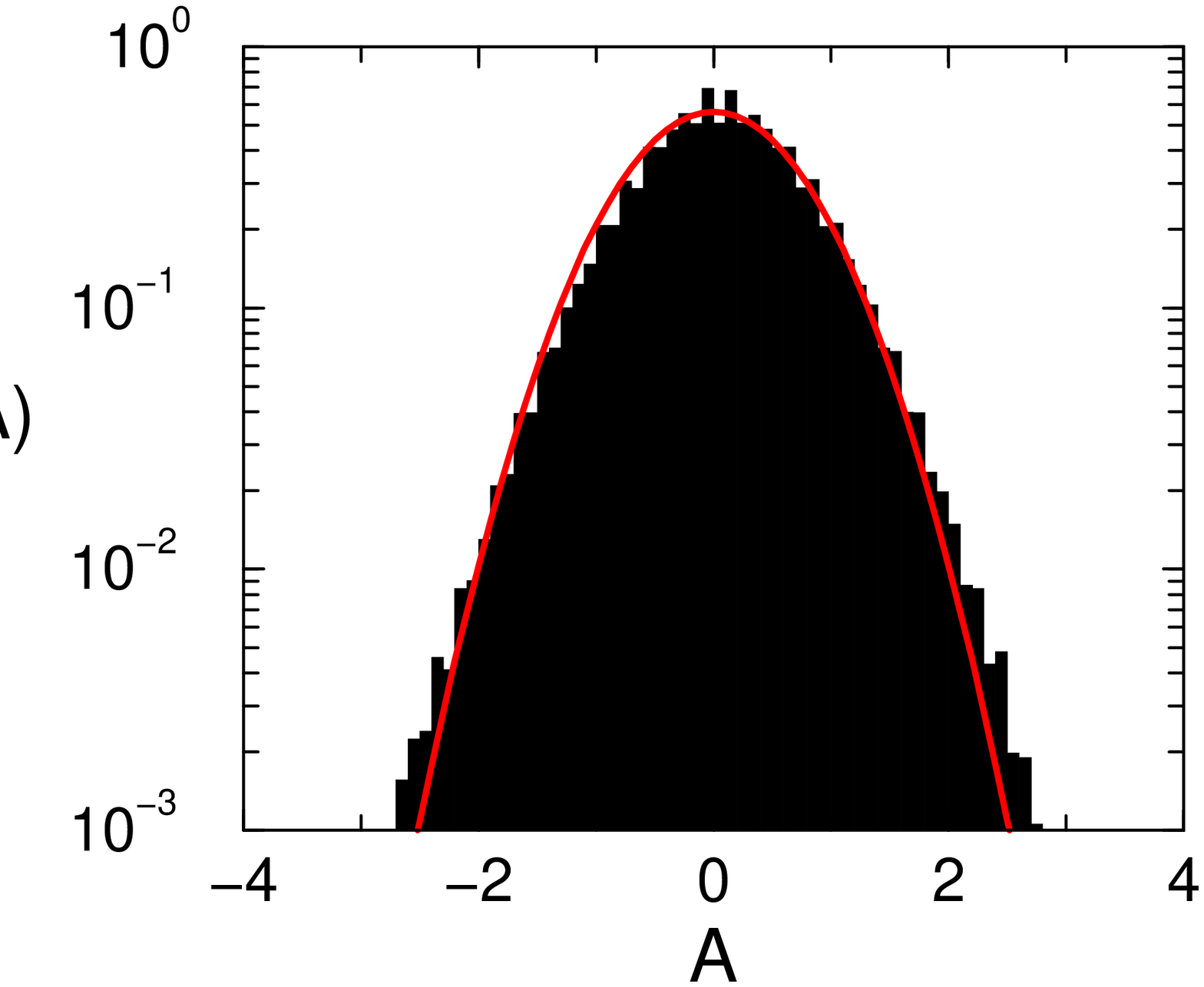} 
\end{center}

\caption{Return distributions of standard MG, left: symmetric phase ($\alpha\approx0.01$), right asymmetric phase ($\alpha\approx 2$). Red line in right panel represents a fit to a Gaussian distribution.\label{fig:mghist}}
\end{figure}

The MG in its most simple setup does not display stylised facts, such
as anomalous fluctuations or temporal correlations of the market
volatility. On the contrary, return distributions are either Gaussian
or of an unrealistic multi-peak shape (due to global oscillations of
the system), see Fig. \ref{fig:mghist}. Within the minimalist approach
of statistical mechanics it is then natural to ask what features need
to be added to produce more realism. It here turns out that an
evolving trading volume appears to be crucial. While in the standard
MG every agent trades precisely one unit at every time to make the
total trading volume equal to the number $N$ of agents, two approaches
have been pursued which allow for dynamically evolving trading
volumes. These are referred to as grand canonical MGs (GCMG)
\cite{gcmg,mgstylised} and MGs with dynamically evolving capitals
\cite{Challet&al} respectively and will be detailed in the following
section. GCMG here refers to MGs in which each agent is given the
option to abstain at any given trading period. If they decide to trade
they still trade one unit, but due to the number of active agents
fluctuating in time, the total trading volume will evolve
accordingly. The second approach consists in MGs with dynamically
evolving capitals, that is the wealth of each agent evolves in time
according to his success or otherwise in the game, and it is assumed
that the amount traded by a given agent is proportional to his wealth
at the time. Since the wealths change over time so does the trading
volume.

We note at this stage that this tracing back of the emergence of
stylised facts to a modulated total trading volume can be criticized,
as liquidity effects normally also need to be taken into account in
real markets. If one assumes that the total trading action $A(t)$ is
related to the evolution of the (logarithmic) price via $\log
p(t+1)=\log p(t)+\frac{A(t)}{\lambda}$, then the role of the liquidity
$\lambda$ needs to be considered carefully. It controls the impact of
the excess demand $A(t)$ on the price returns, and may well vary in
time as well, i.e. $\lambda=\lambda(t)$. We will here disregard this
fact, and will describe the emergence of stylised facts in GCMGs and
MGs with dynamical capitals in the following section, before we then
turn to a discussion of anomalous fluctuations in agent-based models
different from the MG.

\subsection{Anomalous fluctuations in MGs}
\subsubsection{Grand canonical MGs}

In grand-canonical MGs agents are equipped with $S\geq 1$ `active'
strategies, that is strategy tables with binary entries $\{-1,1\}$ for
each history, prescribing to buy or sell. Furthermore they each hold
one null-strategy, i.e. a trivial strategy which prescribes to abstain
for any given history-string ($a_{is}^\mu\equiv 0$). For each of these
$S+1$ strategies they then keep score values as usual and play the one
with the highest score at any time-step. An additional parameter
$\varepsilon$ is here introduced, and plays the role of a disincentive
for the players to be active. This is implemented by subtracting an
amount $\varepsilon$ of the score of any strategy but the zero one at
any time step, so that the score update rules for the `active'
strategies read
\be
U_{is}(t+1)=U_{is}(t)-a_{is}^{\mu(t)}A^{\mu(t)}-\varepsilon.
\ee
 The zero-strategy will always carry score zero. Thus agents trade
 only if the marginal payoff from trading exceeds a threshold
 $\varepsilon$. Modifying $\varepsilon$ thus can be understood as
 introducing a trading fee to the model \cite{BiGaMa}.
\begin{figure}[t]
\vspace*{1mm}
\begin{tabular}{c}
~~~~~~~~~~~~~~~~~~~~~~~~~~~~\epsfxsize=80mm  \epsffile{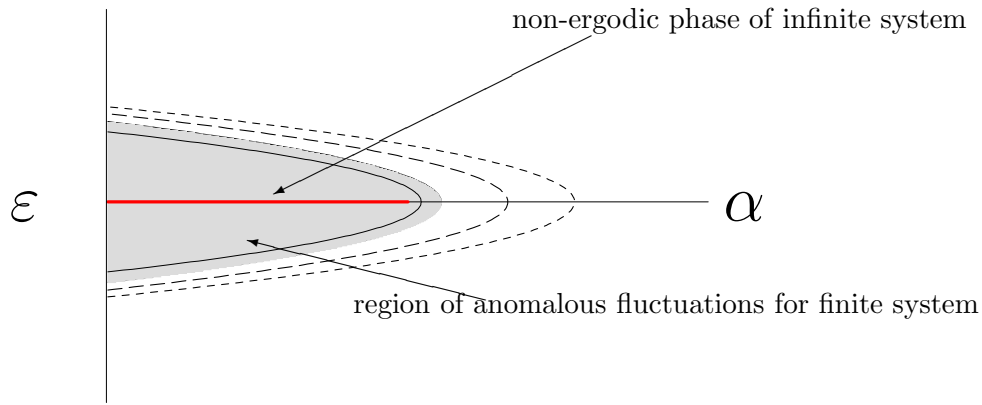} 
\end{tabular}
\put(0,0){\Huge $\alpha$}
\put(-270,0){\Huge $\varepsilon$}
\put(-50,70){\vector(-2,-1){120}}
\put(-80,73){\small non-ergodic phase of infinite system}
\put(-90,-30){\vector(-4,1){90}}
\put(-140,-35){\small region of anomalous fluctuations for finite system} 
\vspace*{4mm} \caption{Phase diagram of the grand-canonical MG. Stylised facts are observed in finite systems in a region around the efficient phase of the infinite-size model.}

\label{fig:pg}
\end{figure}
Since the MG by definition is a negative-sum game (i.e. the sum of all
payoffs is negative), agents loose on average when trading (only a
minority of players wins at every time step). Thus in the long-run all
traders would stop trading in the grand-canonical setup. To maintain
trading activity one usually couples the above group of `speculators'
(that is agents who have the option to abstain) to a background of
so-called `producers'. The latter are endowed with only one non-zero
strategy, and do not have the option not to trade, even if in the long
run they lose money on the MG market. They are considered producers
who make their profits from trading outside the model, and who are
forced to be active on the MG market no matter what
(e.g. internationally operating co-operations who need to be active in
currency markets, but make their profits elsewhere).

The resulting phase diagram of the model at a fixed relative number of
speculators and producers is shown in Fig. \ref{fig:pg} in the
$(\alpha,
\varepsilon)$-plane. For $\varepsilon=0$ one observes the standard
transition of the MG, between an unpredictable phase and a predictable
one. The unpredictable phase (in which $H=0$) is marked by a red line
in Fig. \ref{fig:pg} and extends from some value $\alpha_c$ (which
depends on the composition of the population of agents, i.e. the
relative concentrations of producers and speculators) to smaller
$\alpha$. Here the system is in a non-ergodic state, the so-called
turbulent regime. For $\varepsilon\neq 0$ this transition is absent
and the model market is predictable for all $\alpha$. Generally the
predictability vanishes as $H\sim\varepsilon^2$ as $\varepsilon\to 0$
for $\alpha<\alpha_c$.

\begin{figure}[t]
\vspace*{1mm}
\begin{tabular}{c}
~~~~~~~~~~~~~~~~~~~~~~~~~~~~\epsfxsize=80mm  \epsffile{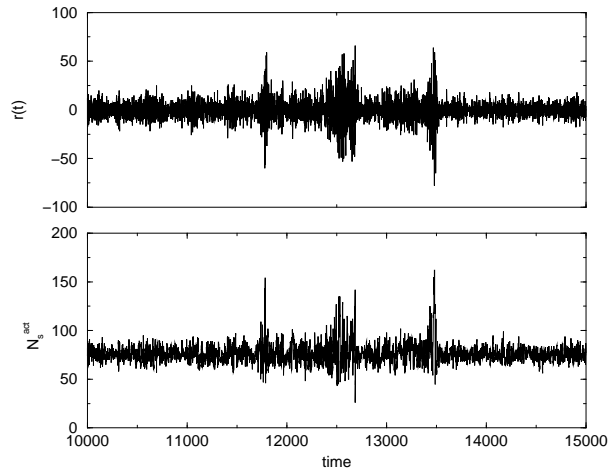} 
\end{tabular}

\vspace*{4mm} \caption{Return time series of a grand-canonical MG in the critical region. From \cite{mgstylised}.}
\label{fig:gcmg}
\end{figure}

The above phase diagram is obtained from the statistical mechanics
theory in the thermodynamic limit $N\to\infty$, and anomalous
behaviour can at most be expected for parameters corresponding to the
red line in Fig. \ref{fig:pg} in the infinite system. Simulations show,
however, that anomalous fluctuations are observed in finite system in a
`critical region' around the critical line, as indicated by the gray
area in Fig. \ref{fig:pg}. Examples of corresponding time-series are
in this critical region are shown in Fig. \ref{fig:gcmg}. As the
system size is increased this critical region shrinks, and finally
reduces to the critical line of the infinite system as the number of
players diverges.

\subsubsection{MGs with evolving capitals}
\begin{figure}[t]
\begin{center}
\includegraphics[angle=270, width=6cm]{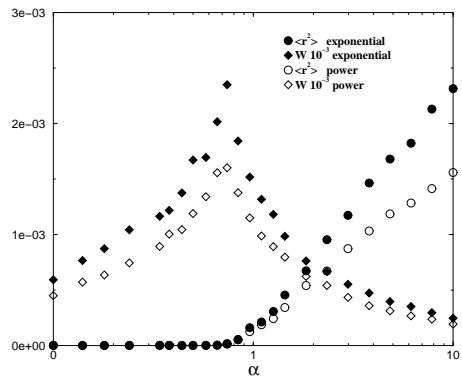}
\end{center} \caption{Return fluctuations and total wealth of the population of traders in an MG with dynamical capitals. From \cite{Challet&al}.}\label{fig:dyncap1}
\end{figure}

In MGs with dynamical capitals each agent $i$ in addition to his $S$
strategies (all different from the null strategy) holds a wealth
$c_i(t)$ which evolves in time depending on his success or otherwise
during trading. It is then assumed that he invests a constant fraction
$\gamma$ of this capital $c_i(t)$ at time $t$. The evolution of the
capital of player $i$ is then given by
\be
c_i(t+1)=c_i(t)-\gamma c_i(t)b_i\frac{A(t)}{V(t)},
\ee
with $V(t)$ the total trading volume, $A(t)$ the excess demand at time
$t$, and $b_i(t)$ player $i$'s trading decision at time $t$. The minus
sign again reflects the minority game payoff. As in the GCMG one
couples the group of speculative traders (whose volume is taken to
evolve in time according to the above rule) to a group of producers,
who each trade a constant volume. This model was studied in
\cite{Challet&al}, and it was seen that the standard transition of the
MG is preserved when introducing dynamical capitals (as illustrated in
Fig. \ref{fig:dyncap1}). At low $\alpha$ the system is in an efficient
phase, which turns out to be an absorbing state of the dynamics. Above
a critical value of $\alpha$ the dynamics do not approach a fixed
point, and one finds $H>0$, just as in the standard MG with fixed
wealth. The overall wealth of the population attains a maximum at
$\alpha_c$ as shown in Fig.  \ref{fig:dyncap1}. We note here that
results of \cite{Challet&al} rely mostly on simulations, as an
analytical solution of MGs with dynamical capitals and $S>1$
strategies per player is tedious due to the presence of fast degrees
of freedom (decisions which strategy to use) as well as slow ones (the
capitals), requiring an adiabatic decoupling. A theory for model with
dynamical capitals and only $S=1$ strategy per player is in progress
\cite{Gallainprogr}, and shows that the standard MG transition is present also in
this simplified system.

\begin{figure}[t]
\vspace*{1mm}
\begin{tabular}{cc}
\epsfxsize=60mm  \epsffile{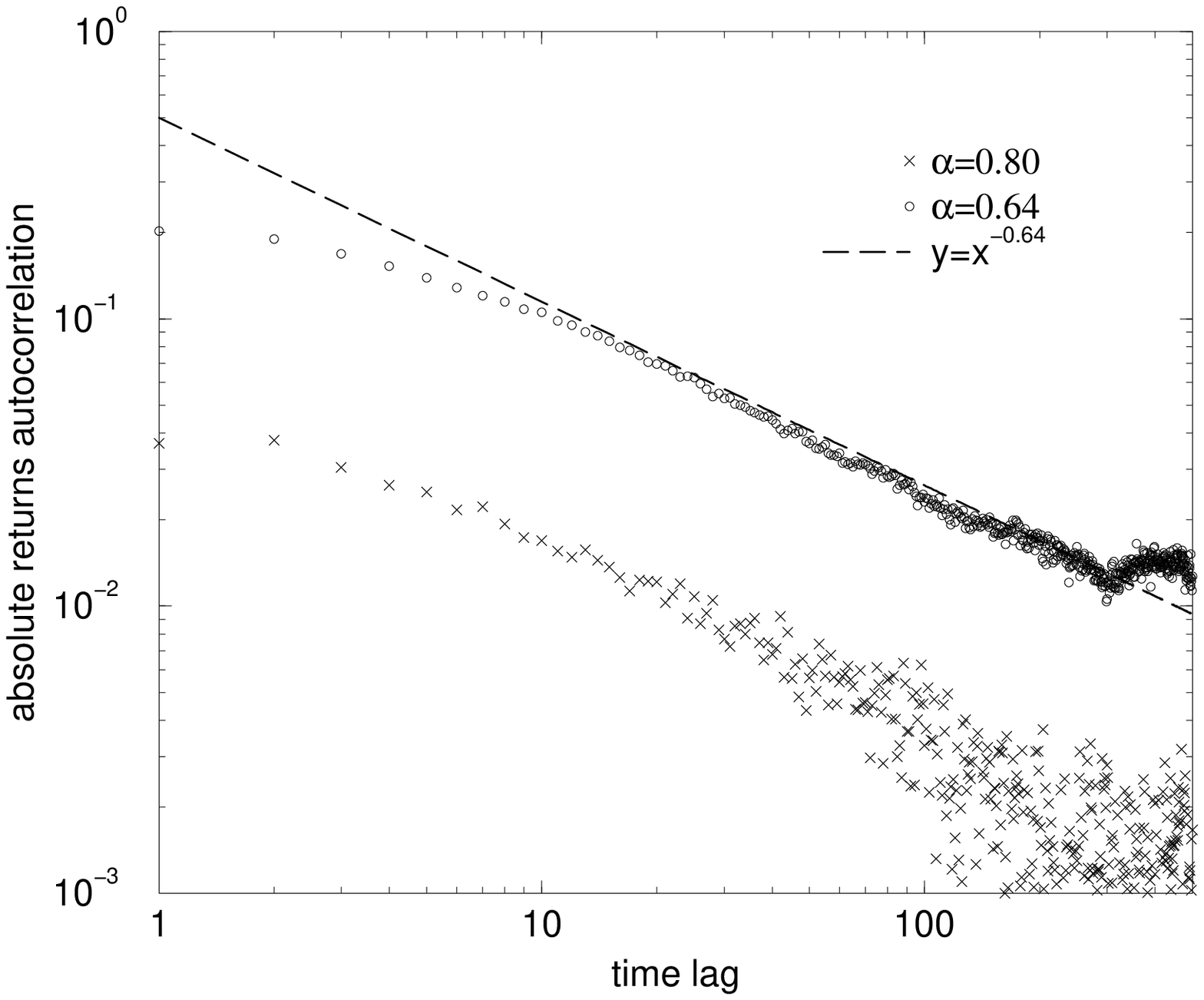}~~~&~~~\epsfxsize=60mm  \epsffile{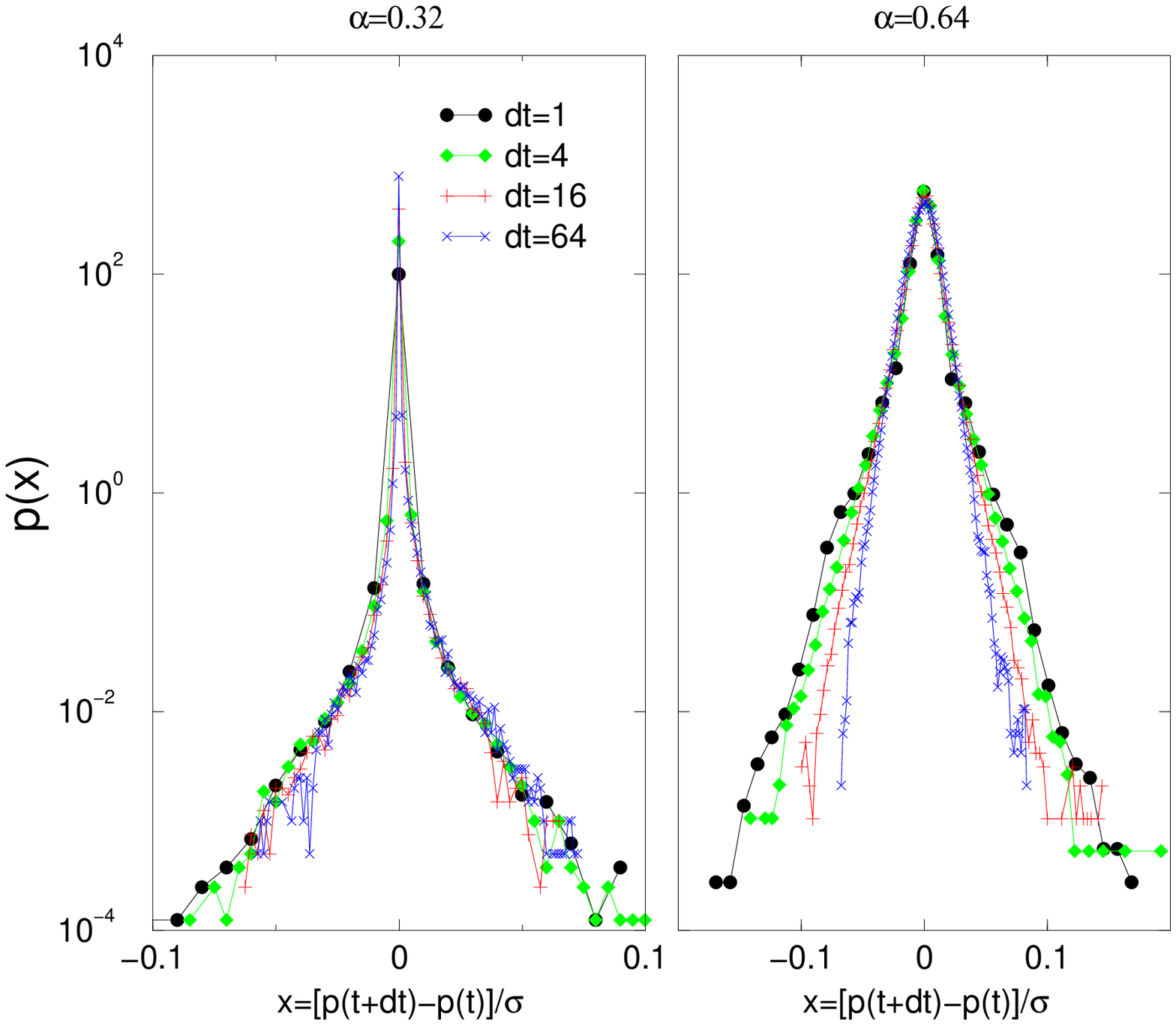} 
\end{tabular}

\vspace*{4mm} \caption{{\bf Left:} Correlation function of absolute returns in an MG with dynamical capitals. The phase transition point is located at $\alpha_c\approx 0.6$ for the parameters used, volatility clustering becomes pronounced close to the transition. {\bf Right:} Distribution of returns in an MG with dynamical capitals. With the parameters used one has $\alpha_c\approx 0.32$. Close to the transition (left) returns collapse on non-Gaussian distributions for different time-lags $dt$, far from the transition (right) a cross-over to a Gaussian occurs. Figures taken from \cite{Challet&al}.}
\label{fig:dyncap2}
\end{figure}

Similar to the observations in the GCMG simulations reveal the
emergence of anomalous fluctuations in finite systems in a region
around $\alpha_c$. This is illustrated in Fig. \ref{fig:dyncap2},
where the autocorrelation of absolute returns is shown and seen to
exhibit algebraic decay, indicating long-range volatility
correlations. At the same time return distributions close to the
critical point are strongly non-Gaussian (see right panel of
Fig. \ref{fig:dyncap2}). While these distributions are essentially
Gaussian far away from the critical point, they attain fatter and
fatter tails as $\alpha_c$ is approached (from above), so that the
kurtosis of these distributions may well be seen as measure of the
distance from the critical point.

\subsection{Other market models}

The MG is only one out of many market agent-based models developed in
recent years, and although this topical review mostly focuses on the
MG, we will provide a brief list of other models, without claiming to
be exhaustive.

We will firstly discuss some close cousins of the MG, then briefly
mention market models which are conceptionally different. We here partly lean on the more extensive reviews \cite{LeBaron2} and \cite{Hommes,LLS}.

 While the MG is a `one-step' game, i.e. a model in which all trading
 action takes place in a single step, this may seem incompatible with
 speculative market which have an intrinsic intertemporal nature:
 buying at a price $p(t)$ is only profitable if one is able to sell at
 a higher price $p(t_1)$ where $t_1>t$. This lead the authors of
 \cite{GiBou,BouGiaMe} as well as Andersen and Sornette
 \cite{dollargame} to use a different payoff function,
\be
g_i(t+1)=a_i(t)A(t+1).
\label{eq:dollargame}
\ee
The model of \cite{dollargame} is here also known as the `dollar
game'. Payoffs of the form of (\ref{eq:dollargame}) imply that
strategies involve two periods of time. The dynamics induced by this
is characterized by different regimes in which the minority and
majority nature of the interaction alternate \cite{FeMa}. However the
majority rule prevails more often and gives rise to a phenomenon quite
similar to bubble phases in real markets. See also \cite{Challet} for
an extension of MGs to multi-step models.

 Agent based models can roughly be divided into two classes: {\em
 few-strategy} and {\em many-strategy} models.  When they first were
 introduced, models of financial markets were intended to recreate the
 situation observed in real markets as closely as possible and for
 this reason they were called `artificial financial
 markets'. Initially the main purpose of these models was to carefully
 analyze a small number of strategies used by agents, which is why
 they are referred to as few-strategies models. Example of such models
 are those by Kirman \cite{Kirman} and by Frankel and Froot \cite{FF}.
 Concepts such as chartist and fundamentalist behaviour were first introduced
  in the context of agent-based simulation in these models.

In many-strategy models, on the other hand, the dynamic ecology of
strategies is studied with the aim of understanding their co-evolution over
time and to see what strategies will survive and which will die. The
Santa Fe Artificial Stock Market (SF-ASM)
\cite{Arthuretal,LBAP} is probably the most known example of
this type of models.  Its main objective is to characterise the
conditions under which the market converges to an equilibrium of
rational expectations.  In the Genoa artificial stock market
\cite{Rabertoetal} randomly chosen traders place a limit buy or sell
order (see also \cite{Bouchaud&Potters} for further details) according
their budget limitation, and they show herding behavior similar to the
one studied in the Cont-Bouchaud model \cite{ContBo}. We would also
like to mention the Levy-Levy-Solomon model \cite{LeLeSo,LLS} in which
agents can switch between a risky asset and a riskless bond and do not
use complicated strategies, but try to maximize a one-period utility
function which implies a wealth-dependent impact of agents on prices,
in contrast with the other models mentioned. Strategy switching,
finally is at the centre of the model by Lux and Marchesi
\cite{LuxMarchesi}. In this model a mixed population of noise traders
(who can either be optimistic or pessimistic), and so-called
fundamentalists is studied. Agents can then switch between these
groups dynamically. Finally, there are of course other market models
which deserve mentioning in principle, but which we can not list here due
to space limitations.

With this plethora of models at hand, all or at least most of which in
some regions of their respective parameter spaces reproduce
market-like stylised facts it is of course legitimate to ask for a
justification for this variety of models, and ultimately for a
selection among them. Apart from its application as a model of a
market, the MG presumably has its own right as a spin glass system and
has introduced new types of complexity and phase transitions to the
theory of disordered systems. Its appeal thus rests, as pointed out
already above, in its position at the boundary of a realistic (within
reason), though analytically solvable complex system.

\section{Towards more realism: heterogeneous populations of agents}
\subsection{General remarks}
\begin{figure}[!]
\begin{center}
\subfigure{\scalebox{.30}{\includegraphics{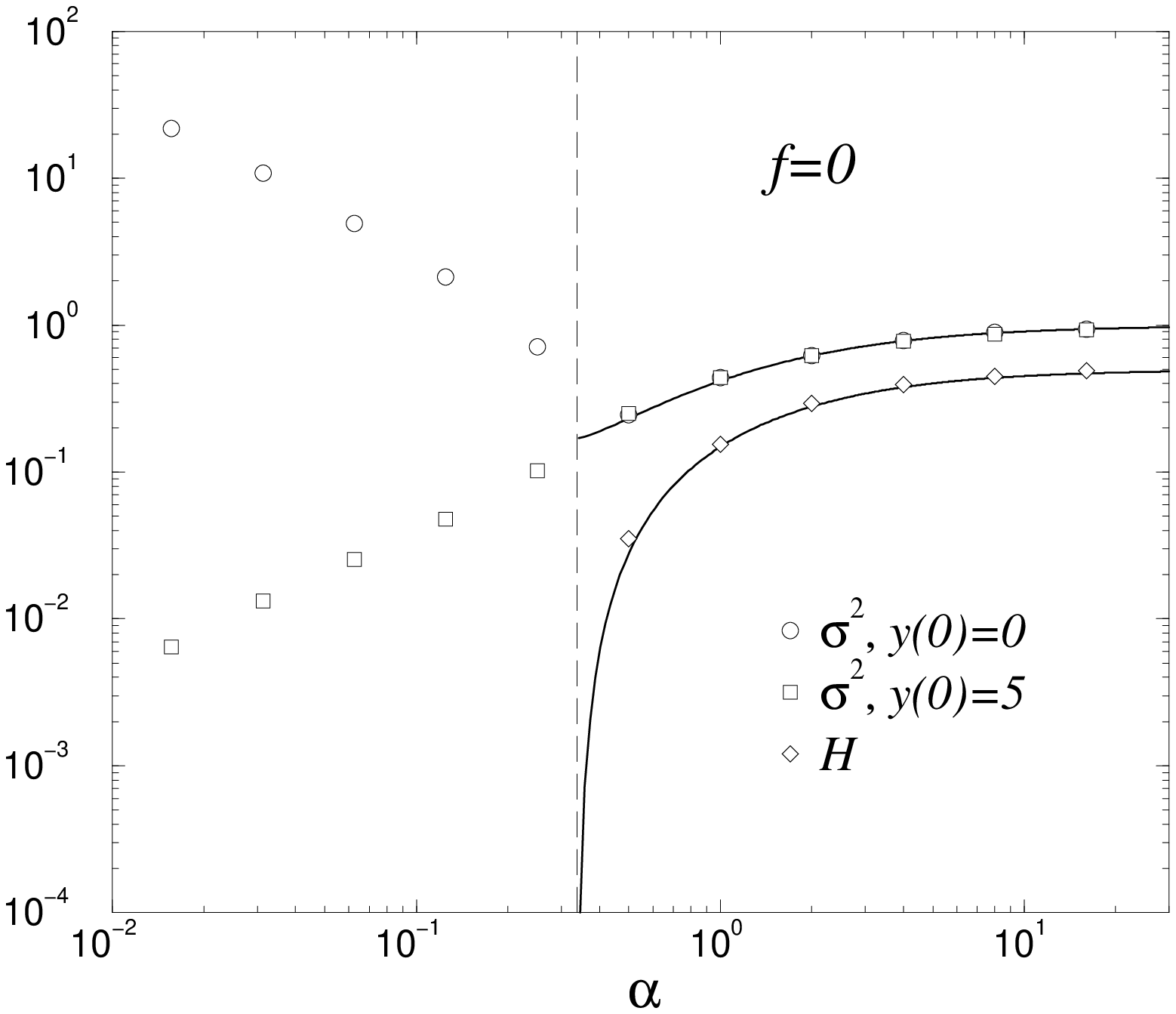}}}
\subfigure{\scalebox{.30}{\includegraphics{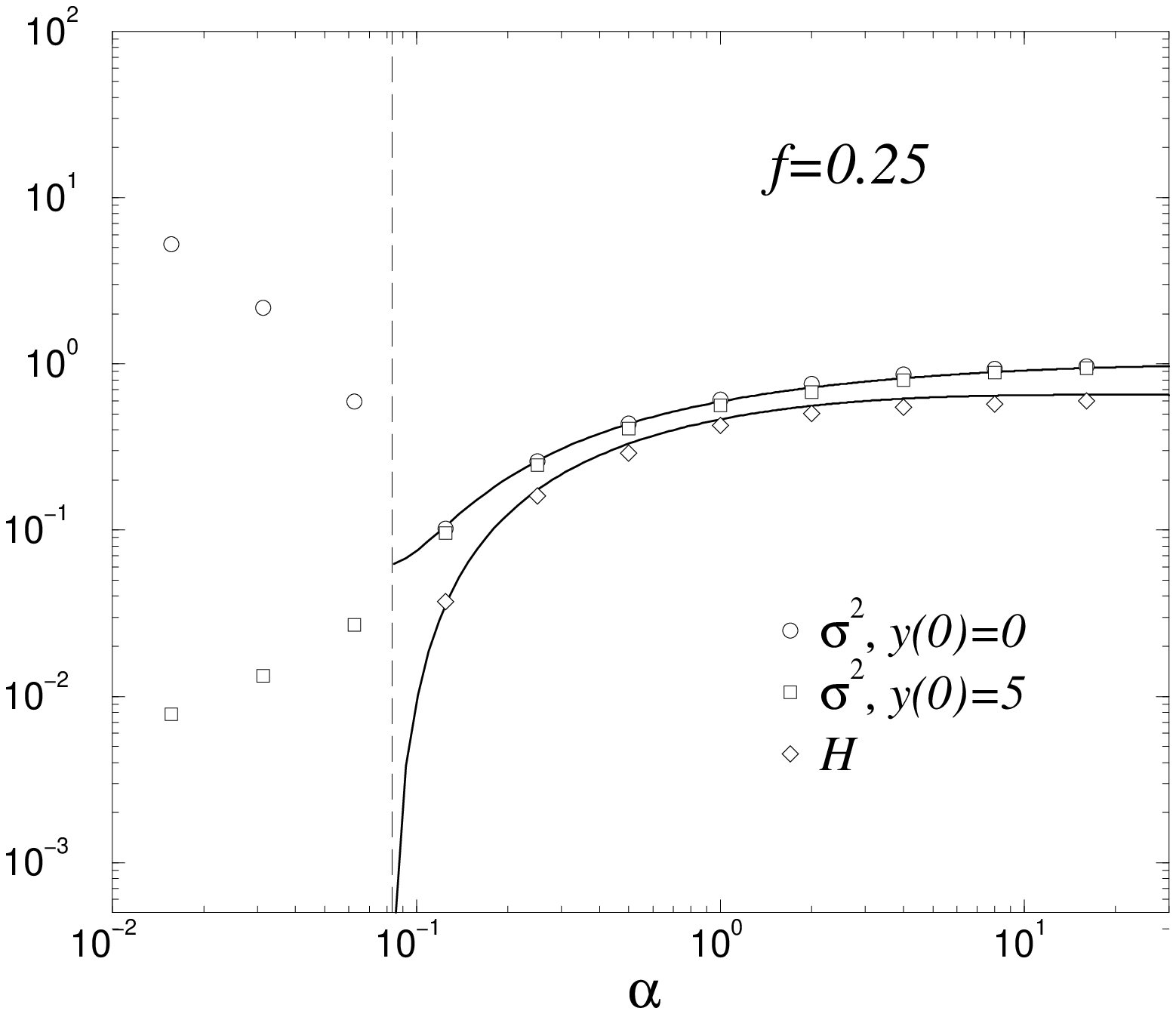}}}\\
\subfigure{\scalebox{.30}{\includegraphics{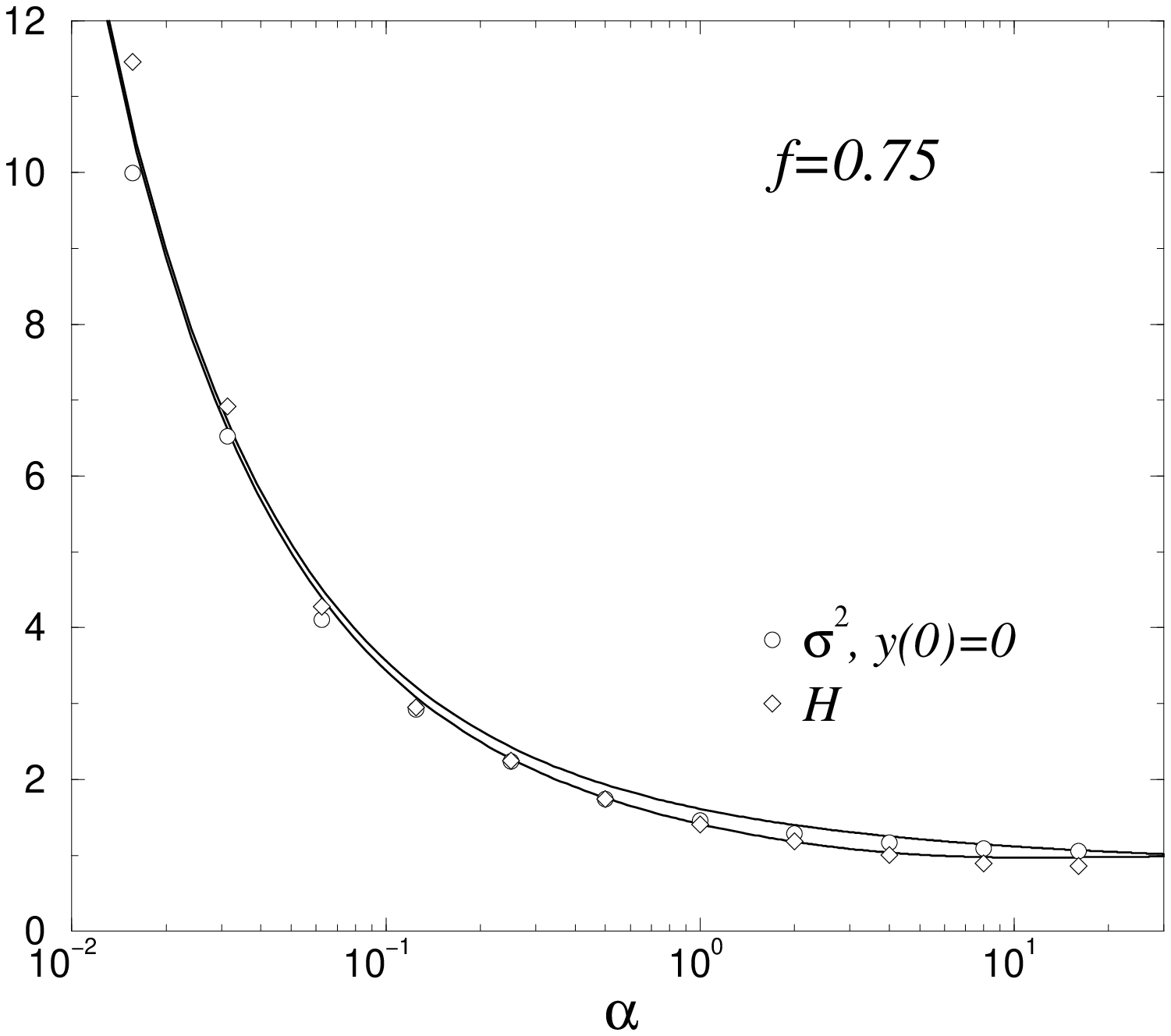}}}
\subfigure{\scalebox{.30}{\includegraphics{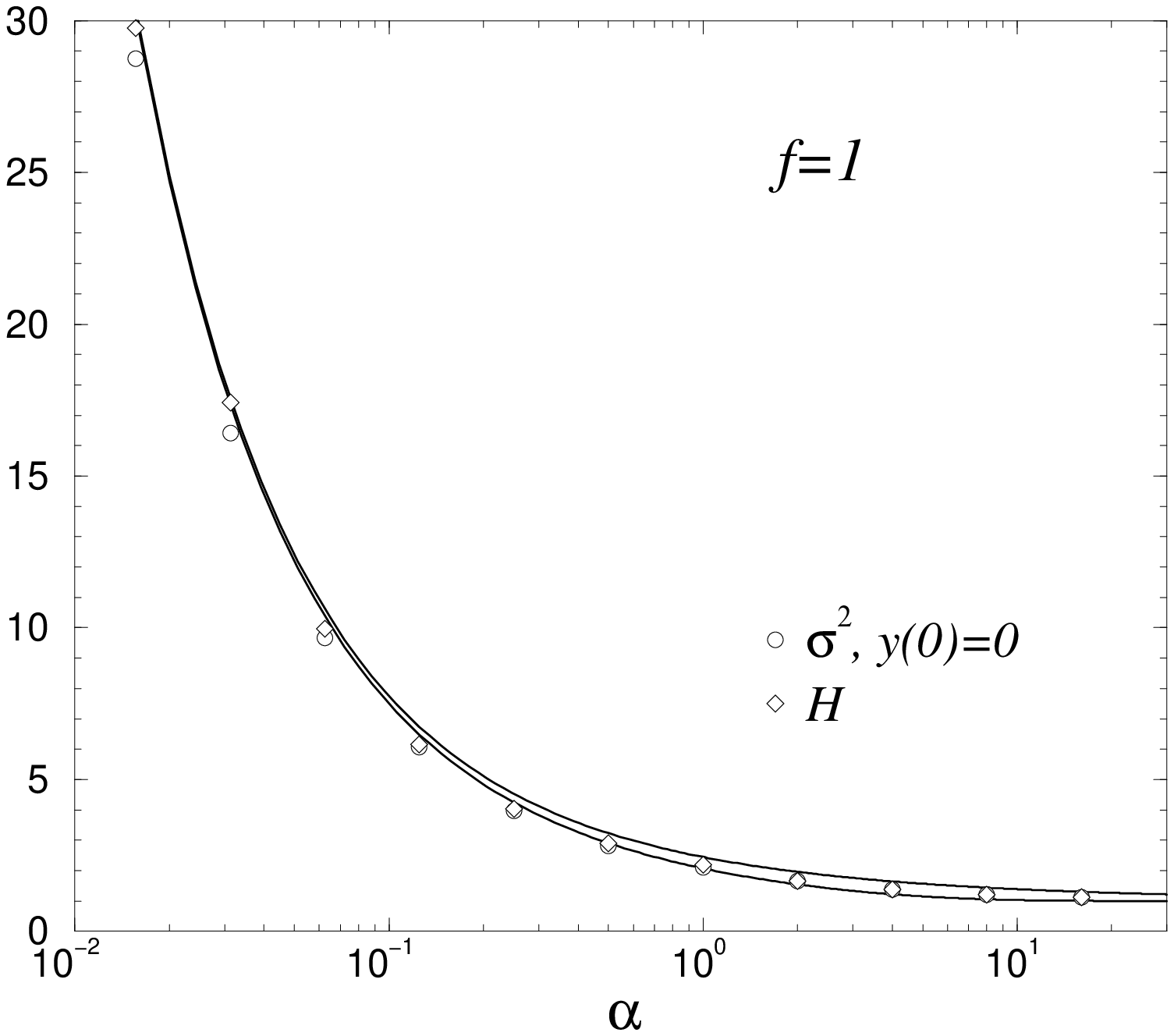}}}
\caption{Behavior of $\sigma^2$ and $H$ vs $\alpha$ in the mixed minority/majority game for different proportions of trend-followers and contrarians. $f$ denotes the fraction of majority games players (trend-followers). From \cite{DeMaGiarMose03}.\label{fig:mixed}}
\end{center}

\end{figure}

One of the main drawbacks of the MG as a market model is its simple
set-up. Apart from the random assignments of strategies there is no
heterogeneity in the different types of agents, whereas real markets
are composed of different groups of traders, with different
objectives, impact on the market, and trading behaviour. The two
extensions mentioned above (GCMG and MGs with dynamical capitals)
alleviate these constraints, but more variety and extensions are
possible and have been studied by various authors over the last
years. We summarise some of them in the following sections.

\subsection{Mixed majority and minority games}
Most notably, extensions towards mixed minority and majority games
have been studied in \cite{DeMaGiarMose03}, in order to emulate mixed
populations of trend-followers and contrarians. As mentioned above, MG
mechanisms can be derived from expectations of agents on the future
price movements, assuming contrarian behaviour. MG agents expect the
price to go down in the future if it went up in the past, and vice
versa. Hence they prefer minority trading. Trend-followers on the
other hand assume that the tendency in price movements will continue,
hence if the price is rising due to positive excess demand, they will
buy (and analogously sell when the price is going down), and behave as
majority game players. Pure majority games are trivial in the sense
that all agents agree to either buy collectively at all times or to
sell, so that all agents are frozen. The model itself is closely
related to the Hopfield model of neural networks, and is interesting
from a statistical mechanics point of view
\cite{KozlMars03}.  A mixed minority/majority game was studied in
\cite{DeMaGiarMose03}, where a population of $fN$ majority game players and $(1-f)N$ minority game players is considered ($0\leq f\leq 1$). Trend followers are here taken to follow a learning rule
\be
U_{i,s}(t+1)=U_{i,s}(t)+a_{i,s}^{\mu(t)}A(t)
\label{eq:payoff_majority}
\ee
where the only difference between (\ref{eq:payoff_majority}) and
(\ref{eq:mgupdate}) is the sign in front of the term
$a_{i,s}^{\mu(t)}A(t)$.  The main results regarding this model are the
findings that the presence of trend-followers (a) injects information
into the system (in particular an informationally efficient phase with
$H=0$ is present only for $f<0.5$), and (b) increases the fluctuations
at low $\alpha$ (see curves for $f>1/2$ in Fig. \ref{fig:mixed}. A
further generalization of the mixed-model can be found in
\cite{DeMGiMaTe, TeDeMaGi} where players have a payoff function for
their strategies given by
\be
U_{i,s}(t+1)=U_{i,s}(t)+a_{i,s}^{\mu(t)}F(A(t))
\ee
with $F(A)=A-\epsilon A^3$ in \cite{DeMGiMaTe} ($\epsilon$ being a
parameter of the model) and $F(A)=A(t)(\chi(|A|<L)-\chi(|A|>L)$ in
\cite{TeDeMaGi} where $\chi(B)=1$ if $B$ is true (and $0$ otherwise)
and $L$ is a parameter of the model. This kind of choice allows a
player to change his behaviour from contrarian to trend follower and
vice versa in time, depending on the current price movement. In some
phases of these models interesting features are observed (e.g. trends,
bubbles and volatility clustering), we point the reader to the
literature for further details.

\subsection{MG as a platform for simulation of a future market}

\begin{figure}[t!]
\centerline{\includegraphics[width=12pc]{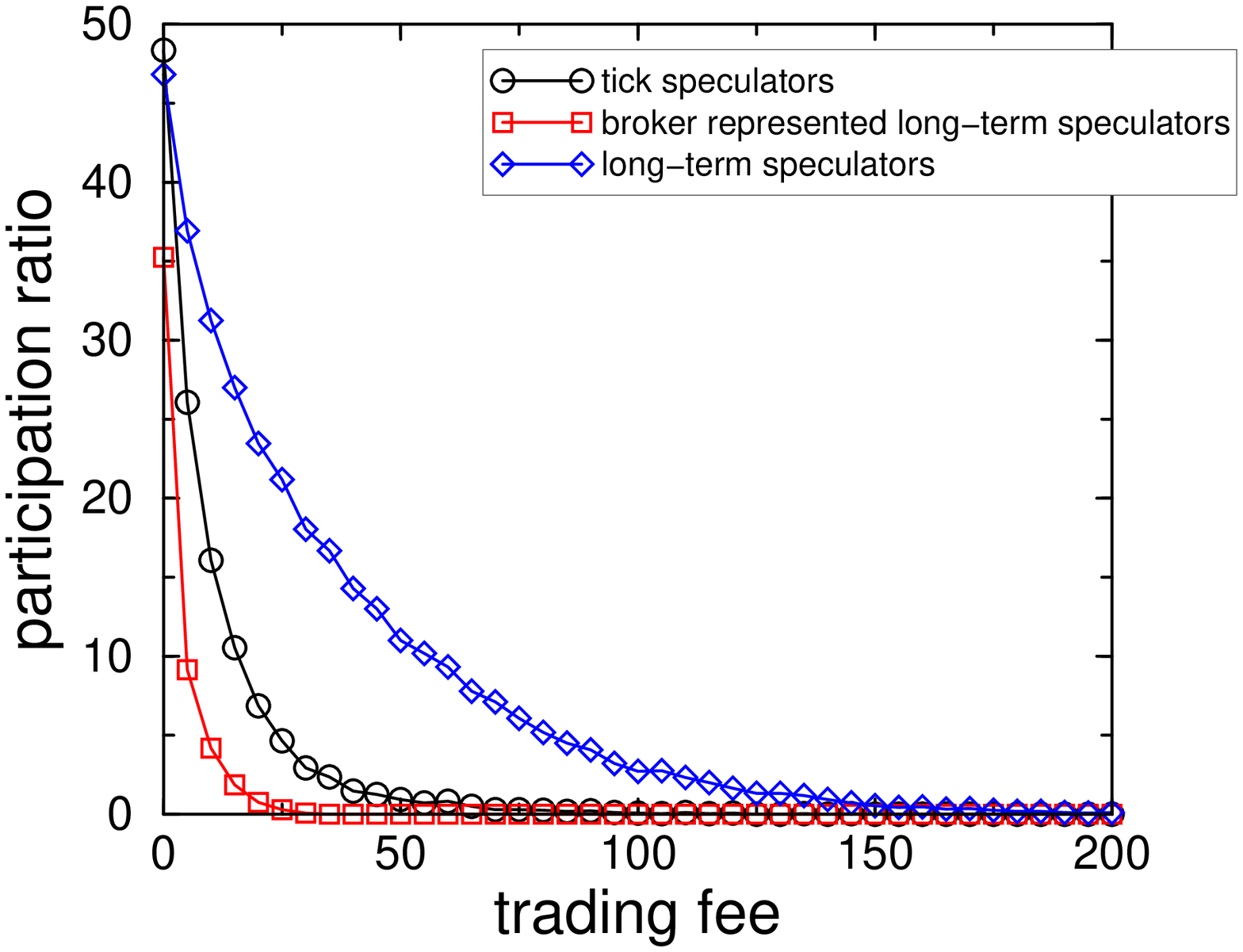}~~~\includegraphics[width=12pc]{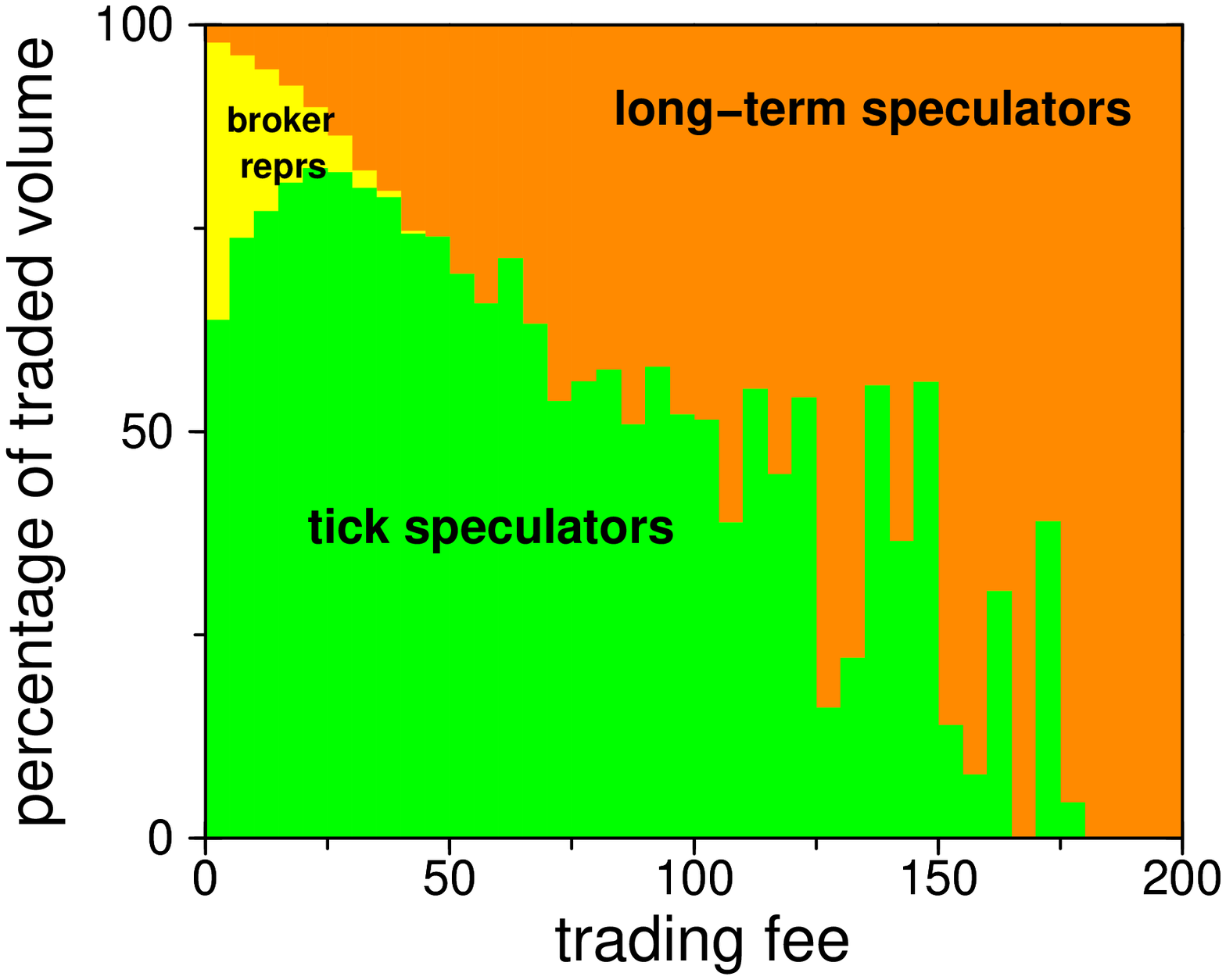}~~~\includegraphics[width=12pc]{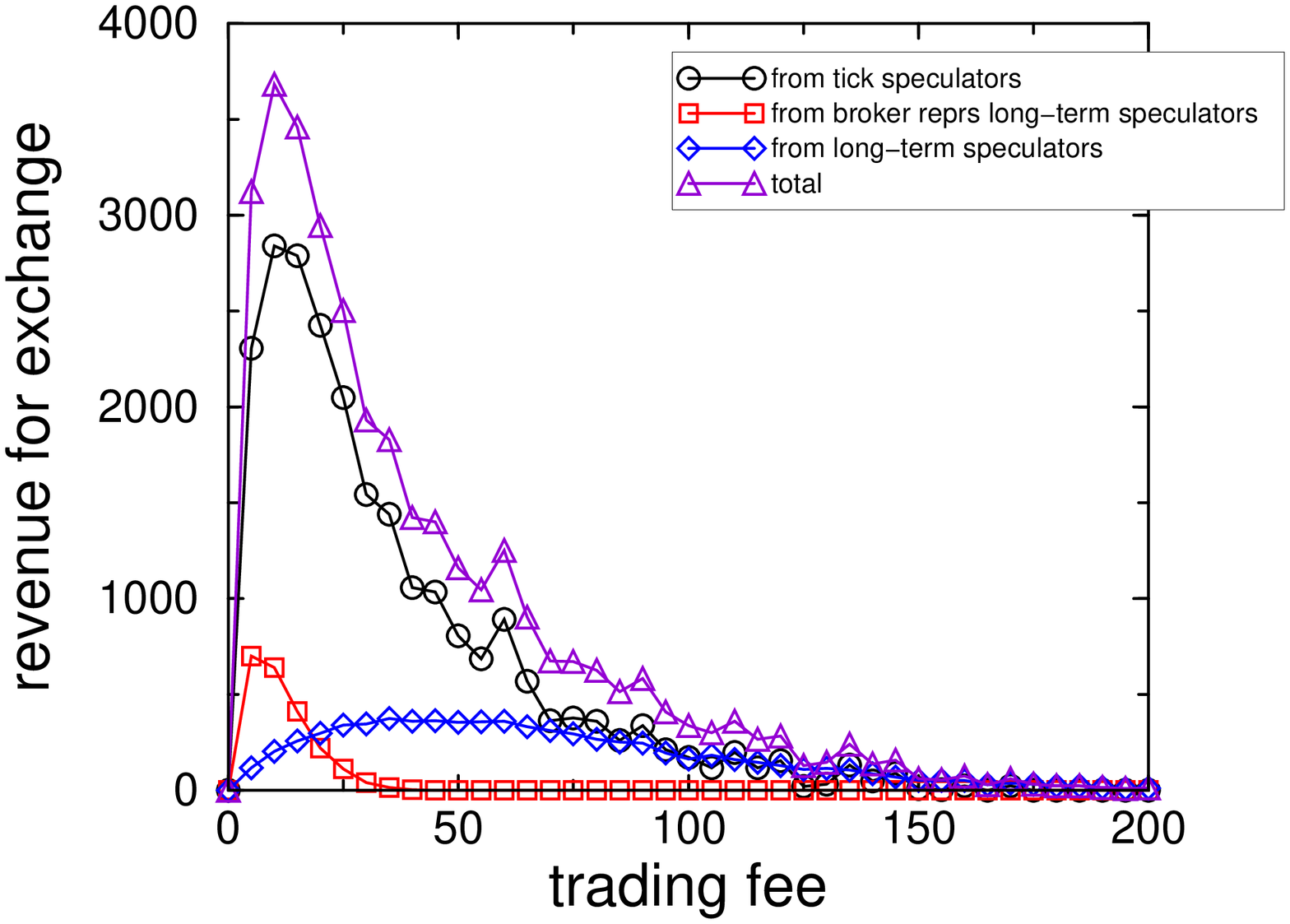}}
\caption{(Colour on-line) {\bf Left:} Participation ratio for tick-traders, long-term speculators and broker represented long-term speculators versus trading fee. {\bf Middle:} Relative percentage of total trading volume attributed to each group. {\bf Right:} revenue for market maker. From \cite{GallaZhang}.\label{fig:ecology}}
\end{figure}

The application of the MG as a platform to study ecologies of traders
can be extended to the simulational study of markets. Current work by
two of the authors \cite{GallaZhang,GallaZhang2} for example uses
MG-based simulations to study the influence of different trading
parameters on the overall performance of different groups of traders
in a market of futures. Large scale simulations with a variety of
traders here allow one to study the interaction between different
groups of traders in detail. In particular diverse populations of
short-term and long-term traders are considered, agents with different
weights are taken into account, and a distinction between direct
traders and broker represented agents can be made.  In
Fig. \ref{fig:ecology} we show results from simulations of a market
with $200$ tick traders (all contrarian), $50$ long-term speculators,
$50$ broker represented long-term speculators and $1000$ broker
represented (background) traders.  While we here depict only results
regarding the influence of a trading fee, the effect of external
market parameters such as so-called `tick size' and `tick value'
parameters and their interdependence can be studied as well. Details
will be reported elsewhere \cite{GallaZhang2}. Note also some recent
work on the effects of Tobin taxes in MG markets
\cite{BiGaMa} in this context.

\section{Use of information in Minority Games and related models}
A further unrealistic simplification is the assumption of uniform available information and equal intellectual capacities of all agents, usually made in simple setups of the MG. More accurate models can here be expected to be more diverse and should account for heterogeneity in these aspects. In particular issues of individually different memory capacities and access to information are here to be considered, the latter leading to models with so-called `private information'.

\subsection{An agent-based model with private information}
A model with private information has been considered in \cite{Berg&al}
and \cite{DeGa}. Although the model studied there is not directly
derived from the MG, it shows similar features. The model is based on
the Shapley-Shubik \cite{ShSh} model of markets, and considers
limitation to private information explicitly. As in the MG one assumes
the existence of a global state of the market $\omega(t)$, which takes
one of $P=\alpha N$ values at each time $t$ (with $N$ the number of
traders in the market), but that agents have access to this
information only through individual information filters $k_i^\omega$,
where $i$ labels the individual agents. On the basis of the external
signal $\omega$ agent $i$ receives `filtered' information $k_i^\omega$
which may hence vary across the population of agents. In the simple
setup of
\cite{Berg&al} and \cite{DeGa} the $\{k_i^\omega\}$ take only binary
values $k_i^\omega\in\{-1,1\}$. The state $\omega$ is also assumed to
determine the (quenched random) return $R^\omega$ of an underlying
asset traded on the model market. Thus the vector
$\mathbf{k}_i=(k_i^1,\dots,k_i^P)$ crucially constrains the ability of
agent $i$ to resolve different states of the market. If for example
$k_i^\omega=-1$ for all $\omega$ then agent $i$ is completely blind
with respect to the actual state of the market, he receives
information $-1$ no matter what the actual value of $\omega(t)$ and
hence cannot distinguish any two different information patterns. If
$k_i^\omega$ is highly correlated with $R^\omega$ (e.g $k_i^\omega=1$
whenever the return $R^\omega$ is positive, and $k_i^\omega=-1$ for
all $\omega$ for which $R^\omega$ is negative), then agent $i$ is very
well able to distinguish states of the market, and to make accurate
predictions on the further price movements.

\begin{figure}[t]
\begin{center}
\includegraphics[width=5cm]{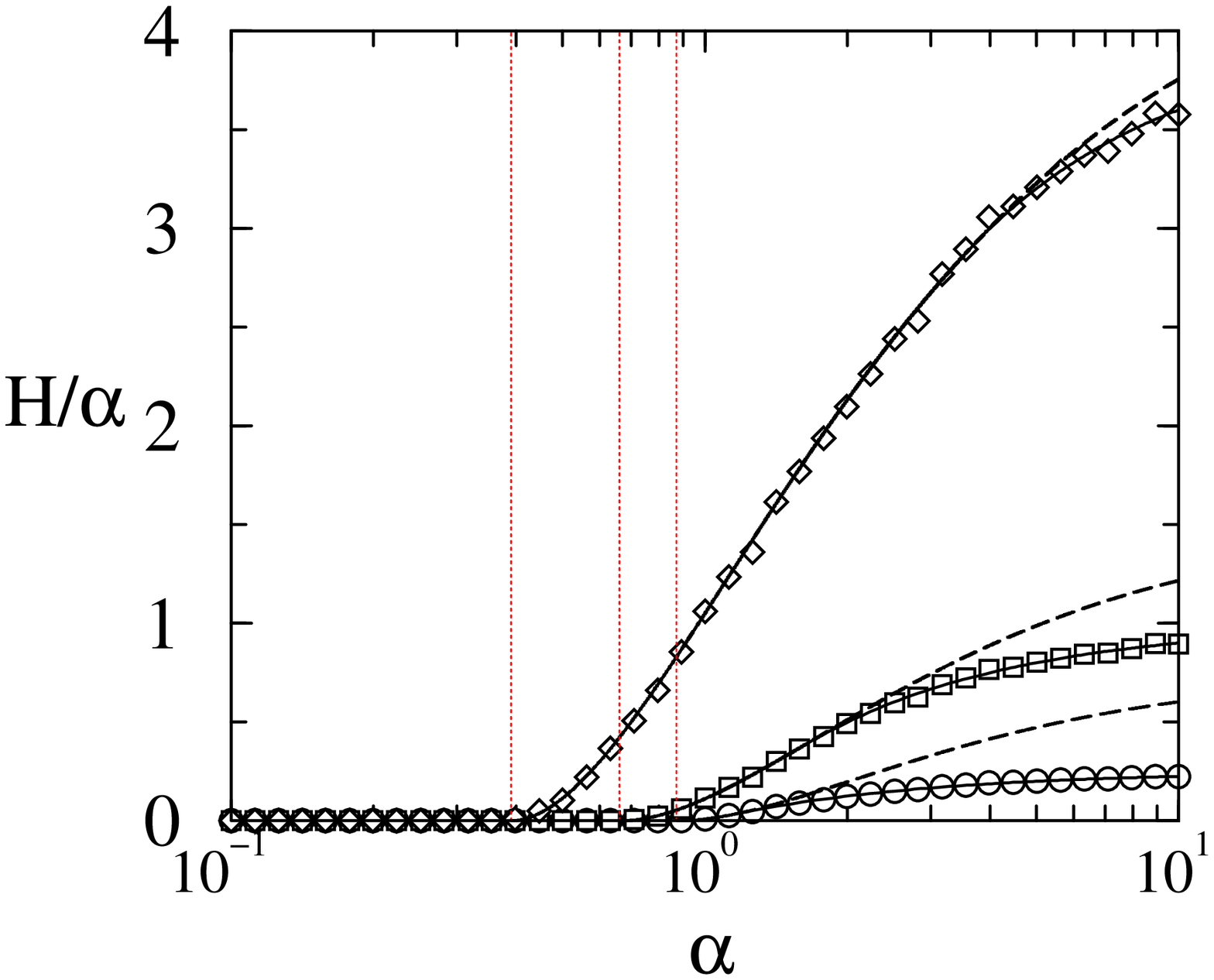}~~~\includegraphics[width=5cm]{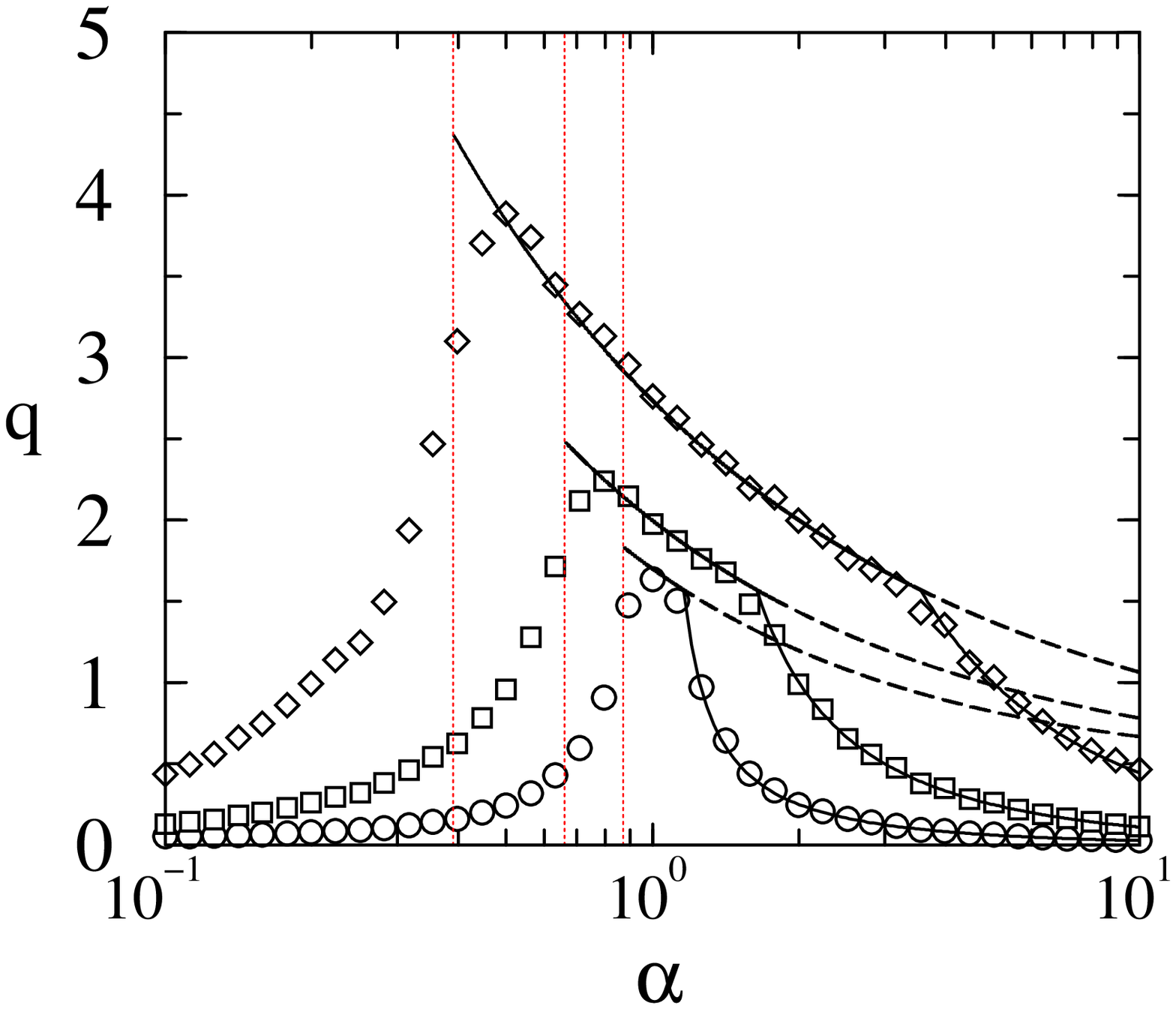}
\end{center}\caption{Predictability $H$ and use of information $q$ in a market model with private information. From \cite{DeGa}.\label{fig:privateinfo1}}
\end{figure}

A detailed statistical mechanics analysis shows that the phase
transition of the MG market model is present also in this context, see
Fig. \ref{fig:privateinfo1}. There we plot the resulting
predictability $H$ as a function of $\alpha$ for different choices of
other model parameters (not specified here), note in particular that
the model can again be solved exactly (solid lines in the figure). The
right panel of Fig. \ref{fig:privateinfo1} depicts an order parameter
$q$ which measures the degree to which agents use their privately
obtained information. At the phase transition $q$ is maximal, and the
agents' decisions crucially depend on the information they receive. At
large or low values of $\alpha$ much less use of the available
information is made, and agents essentially trade independently of the
received binary private information.

\subsection{MG with dynamics on heterogeneous time-scales}
\subsubsection{Heterogeneous time-scales}

\begin{figure}[t]
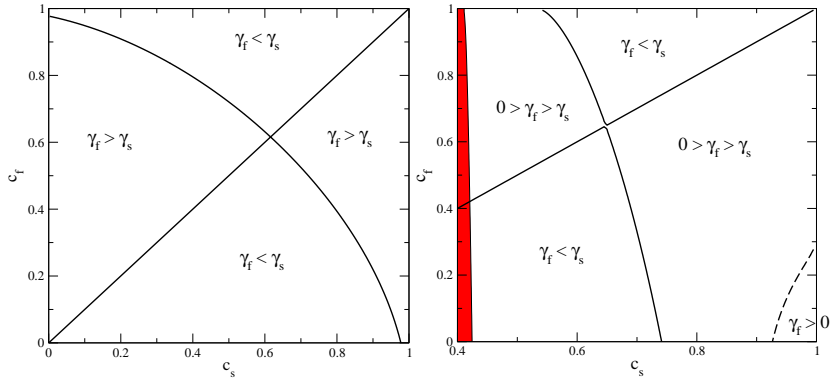

\centerline{\includegraphics*[height=5cm]{contour-a2phi0.5.eps}\includegraphics*[height =5cm]{contour-a0.4phi0.01.eps}}
\caption{Regions of relative advantage for $\phi_f=0.5$ (left graph) and $\phi_f=0.01$ (right graph); the region filled corresponds to the symmetric phase, in which the replica calculus is not valid. From \cite{MoChZh}.}
\label{fig:gain-regions}
\end{figure}

Agents in real markets can differ in time horizon on which they act
and on which they develop their own strategy. This point is often
neglected, although there is of course some work stressing this
particular aspect of real markets \cite{Olsen}. Very recently LeBaron
pointed out that different time scales can be responsible for the
appearance of different beliefs on the market and the co-existence of
these beliefs helps to generate features across many time scales
through symbiotic effects \cite{LeBaron}. In MGs there are several
ways in which one can implement different time scales among agents,
for example through individual learning rates, score memories or
strategy correlation
\cite{MoChZh}. In 2-strategy MGs one can introduce correlation between the strategies of any given agent by the drawing the strategy tables such that
\be
P(a^\mu_{i,1}=a^\mu_{i,-1})=c.
\ee 
In this setup, with probability $c$, the two strategies of any given
player will prescribe the same action as response to the appearance of
history $\mu$. Thus, if $c=0$ player $i$'s strategies are fully
anti-correlated, $a_{i,1}^\mu=-a_{i,-1}^\mu$ for all $\mu$. For
$c=1/2$ one recovers the MG with uncorrelated strategies. For $c=1$
each players holds two fully correlated (i.e. identical strategies),
and is an infinitely `slow' player in the sense that his reaction to a
certain information pattern does not change over time.  The smaller
$c$ the faster are the agents, at least in the simplified picture of
this model. In this setup, one is particularly interested in the
information ecology of the model, i.e. to understand whether groups of
different strategy correlation exploit each other, and can live in
symbiosis.  If two different groups, say fast and slow, are
introduced, with $c_f<c_s$ where $f,s$ stand for fast and slow, the
behaviour is already quite rich and can be studied exactly by means of
replica theory. Fig.
\ref{fig:gain-regions} shows that gains $\gamma_f,
\gamma_s$ of fast and slow agents respectively, depend in a highly non-trivial way on the fraction $\phi_f$ of fast players and the correlation parameters $c_f, c_s$ .

Another way to investigate the role of time scales in MGs is to study
the game with just one information pattern (i.e. the limit $P\to 1$)
\cite{Marsili,MC01}. This limiting case is particularly
straightforward to understand analytically. Given $A(t)$ each agent
$i$ receives a payoff $-a_i(t)A(t)$, and keeps a score $\Delta_i(t)$ as follows
\be
\Delta_i(t+1)=\Delta_i(t)-\frac{A(t)}{N},
\ee
and his trading action is then determined by the stochastic rule
\be
P[a_i(t+1)=1]=\frac{1+\Gamma_i\Delta_i(t)}{2}
\ee
with $\Gamma_i$ a learning rate. Thus high values of $\Delta_i(t)$
favour positive trading action $a_i(t+1)=1$ by player $i$. It is then
possible to consider $G$ groups of $\phi_g N$ agents respectively
(with $g=1,\dots,G$ and $\sum_{g=1}^N \phi_g=1$) with all the agents
belonging group $g$ following a learning rule of rate $\Gamma_g$.
Without going into any detail here, we will just point out that the
relative gains of the respective groups can be obtained as
\be
\gamma_g-\gamma= \frac{\Gamma-\Gamma_g}{2(1+\Gamma)}
\label{eq:learning_rates}
\ee
(within some approximation) where $\gamma_g$ is the average gain of
players in group $g$, $\gamma$ the gain averaged over all groups, and
$\Gamma$ the mean learning rate (over all groups). It is interesting
to observe that the smaller $\Gamma_g$ the smaller the loss of group
$g$.  For details and other extensions see \cite{MoChZh}.

\subsubsection{Timing of adaptation}
\begin{figure}[t]
\begin{center}
\includegraphics[width=5cm]{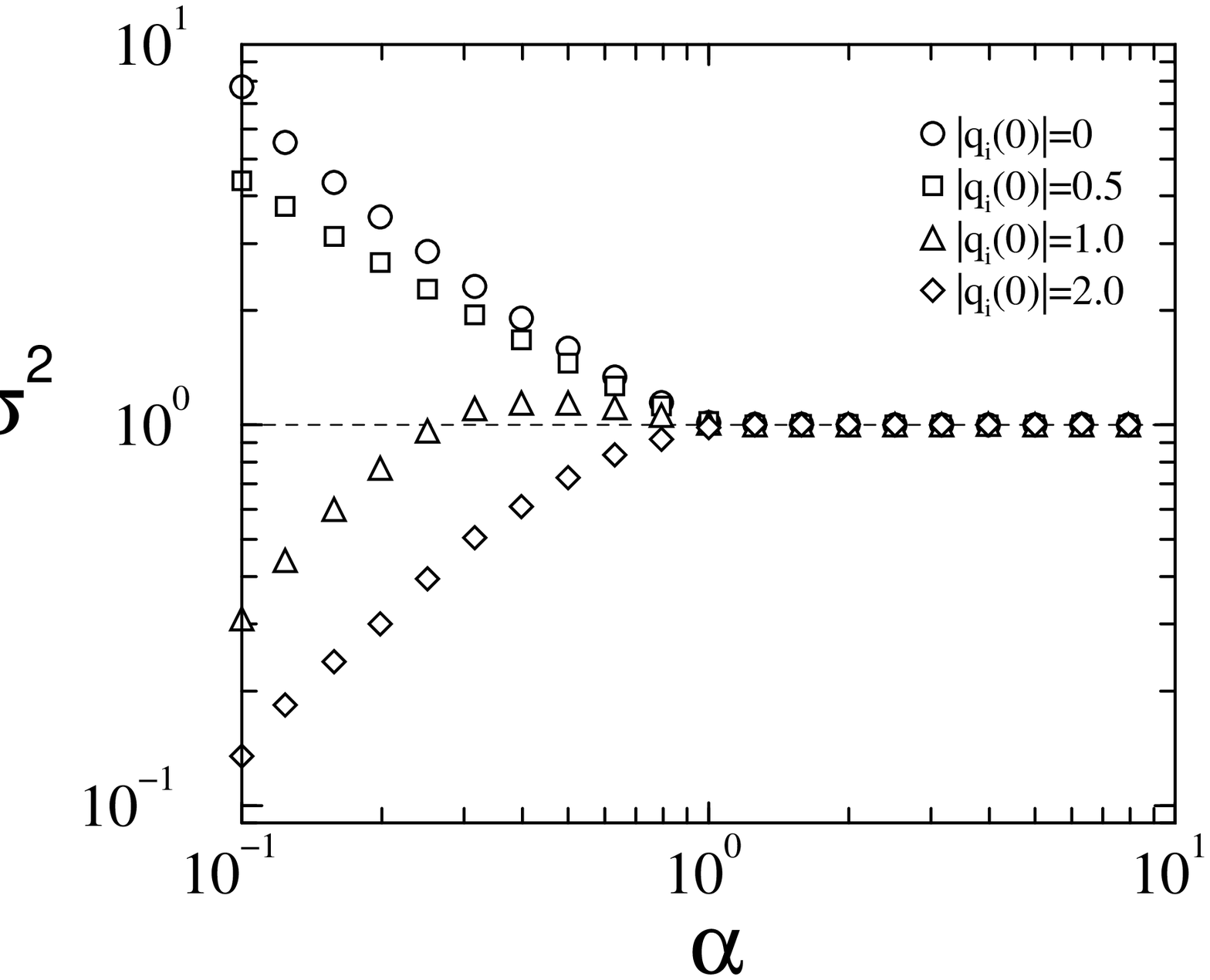} ~~~\includegraphics[width=5cm]{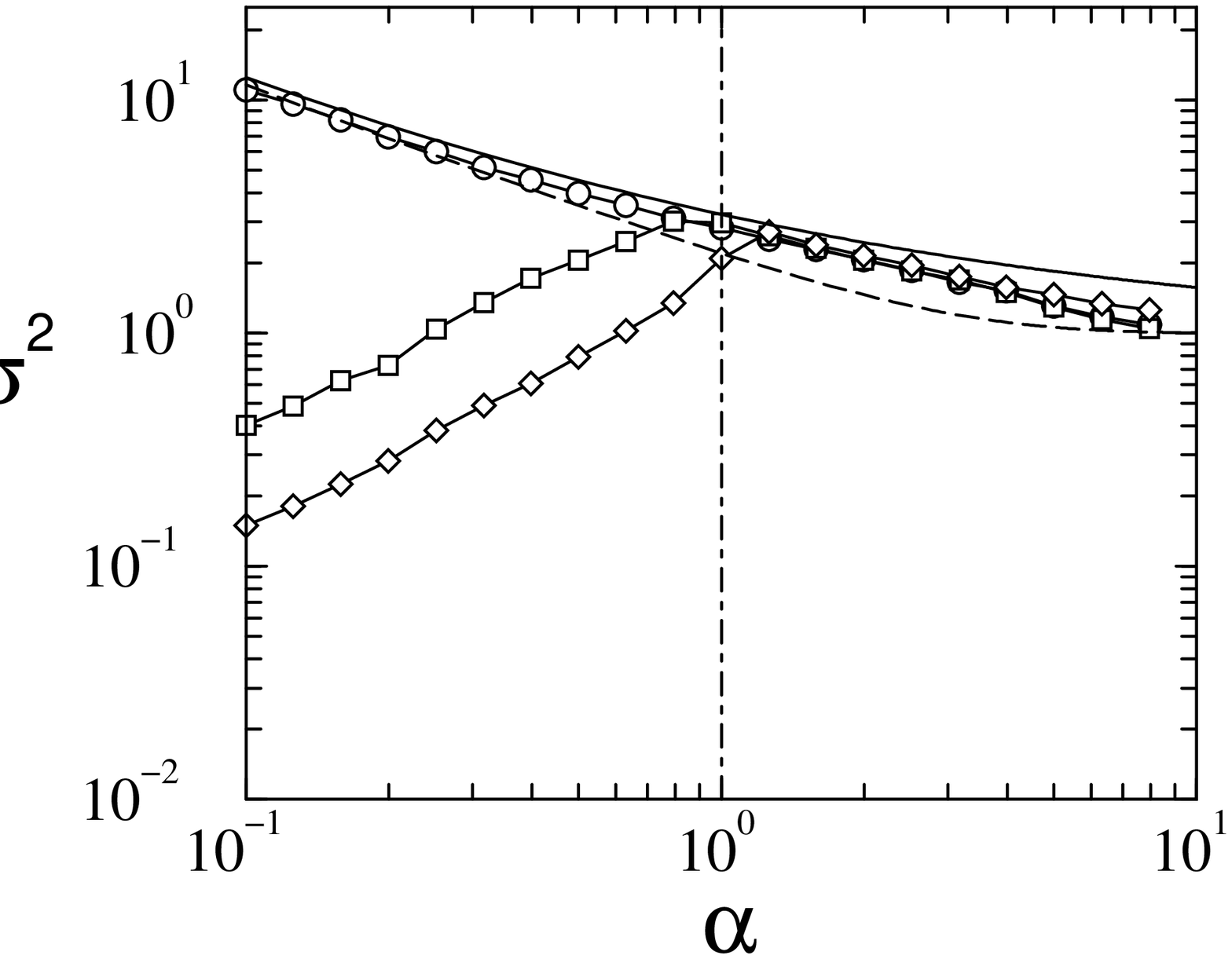}
\end{center}\caption{On-line versus batch game for fully anti-correlated strategies. {\bf Left:} on-line game, started from different initial conditions. {\bf Right:} batch game. Note the difference in the qualitative behaviour. From \cite{SherGall05}.}\label{fig:batchonl}
\end{figure}

\begin{figure}[t]
\begin{center}
\includegraphics[width=6cm]{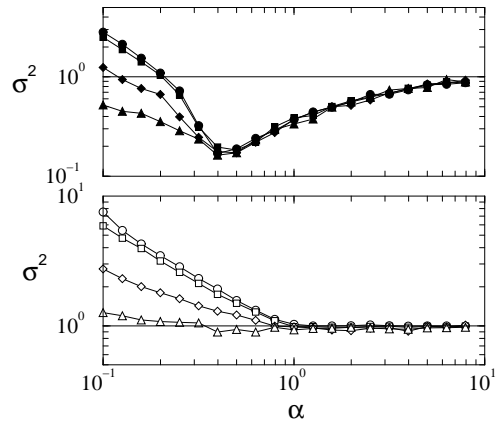} 
\end{center}\caption{Volatility for model with random updates, probability to update $1/M$. Top panel corresponds to MG with uncorrelated strategy assignments, lower panel to the fully anti-correlated case. From \cite{SherGall05}.}\label{fig:1_m}
\end{figure}

The effects of the timing of adaptation on MGs have been studied for
example in
\cite{SherGall03,SherGall05}. One here distinguishes between so-called
`on-line' and `batch' adaptation of the MG agents. In the more usual
`on-line' games, agents adapt to the external flow of information
instantly, and update their strategy scores at every round of the
game, and then choose the trading strategy to use in the next step. In
batch games, agents adapt only after a large number of rounds has been
played, and the corresponding information regarding the performance of
the strategies has been accumulated. An interpolation between both
cases is possible. Specifically, one allows the agents to update their
scores (and re-adapt their strategy choices) only once every $M$
rounds (with accumulative increments over the past $M$ rounds), i.e.
\be
U_{i,s}(t+M)=U_{i,s}(t)-\frac{1}{M}\sum_{\ell=0}^{M-1}a_{i,s}^{\mu(\ell)}A^{\mu(\ell)}[\{U(t)\}].\label{eq:emm}
\ee
 Their choice of strategy then remains
fixed in between two updates (as indicated by the dependence of the
total action on the scores at time $t$ in Eq. (\ref{eq:emm})). The
limit $M\to 1$ reproduces the on-line case, the case $M\gg P$ (with $P$
the number of possible values of the information) is referred to as
the `batch' game, in which an effective average over all information
patterns is performed. While the volatility of standard MGs with
uncorrelated strategies is not sensitive to the choice of on-line
versus batch dynamics, qualitative differences are found in MGs with
fully anti-correlated strategy assignments ($c=0$ in the notation of the previous section), see
Fig. \ref{fig:batchonl}. Adaptation at randomly chosen time-intervals
of mean $M$ finally reduces the volatility in the efficient phase of the
market, but not above the phase transition, see Fig. \ref{fig:1_m}.

\subsection{Cost of information}
Ideally one would like to consider models in which agents can acquire
information at some cost, i.e in which they pay for the use of
expertise. To the knowledge of the authors, no such attempts have yet
been made in an analytical statistical mechanics approach. One may
also consider models in which agents have both public and private
information at their disposal, and have to decide which to use. One
may think of globally available information through newspapers and
other media, versus the recommendations of a small circle of private
advisors and/or friends. 

\section{Summary and outlook on future work}
In summary we have reviewed the recent progress in the description of
financial markets through simple agent-based models, accessible by the
techniques of theoretical statistical mechanics. The MG in particular
can be seen as a minimalist market model, which can be treated
analytically in its basic setup and is now considered to be
essentially fully understood. While in its original setup of each
agent trading one unit at any given time step the MG does not display
anomalous fluctuations and stylised facts as seen in real market data,
only minor modifications are necessary to make the model more
realistic in this sense, without giving up the analytical
tractability. MGs with a finite number of agents and with dynamical
capitals and/or with an option of the agents to abstain have been
shown to display anomalous fluctuations close to their phase
transitions, similarly to systems of statistical mechanics exhibiting
large scale correlations and self-similarity near their critical
points. These observations call for further analysis of such variants
of the MG, for example by means of renormalisation group techniques.

The MG can also serve as a platform for more diverse market
simulations, and an ecology of market participants can be studied upon
introducing diversified types of players in the MG. These may include
contrarians and trend-followers respectively, as well as agents
trading on different time-scales and/or with different trading
volumes, and agents of different memory capacities. While it is
presumably unlikely that real market trading decisions can be taken on
the basis of MG simulations, the model as such and its diverse
variants and extensions offer a promising approach to gain a better
understanding of the interplay of market parameters and the ecology of
populations of traders, based on models at the boundary of analytical
solvability. Future research in this direction may thus be of interest
both in an academic environment as well as for practitioners.

\section*{Acknowledgements}
This work was supported by the European Community's Human Potential
Programme under contract HPRN-CT-2002-00319, STIPCO and by EVERGROW,
integrated project No. 1935 in the complex systems initiative of the
Future and Emerging Technologies directorate of the IST Priority, EU
Sixth Framework. The authors would like to acknowledge collaboration
with Damien Challet, Ton Coolen, Andrea De Martino, Irene Giardina,
Matteo Marsili and David Sherrington on some of the material reviewed here.

\end{document}